\documentclass[aps,pra,twocolumn,eqsecnum,superscriptaddress,
showpacs,floatfix,nofootinbib]{revtex4}

\usepackage{psfrag,graphicx} \usepackage{dcolumn}
\usepackage{amsmath,amssymb} \usepackage{bm}
\usepackage[dvips]{epsfig,color} \usepackage[english]{babel}

 \newcommand{\ra}{\rangle}

\newcommand{\proj}[1]{\mbox{$|#1\rangle \!\langle #1 |$}}
\newcommand{\ev}[1]{\mbox{$\langle #1 \rangle$}}

 \def\be{\begin{equation}} \def\ee{\end{equation}}
\def\bea{\begin{eqnarray}} \def\eea{\end{eqnarray}}
 
\def\H{{\cal H}}

\def\ch{\raisebox{0.3ex}{$\chi$}} 
\def\tr{ \mbox{tr}} \def\tr{{\rm tr}} 
\def\1/2{\frac{1}{2}} 
\def\Ma{\check{a}}
\def\Mb{\check{b}}
\def\Mc{\check{c}}
\def\Md{\check{d}}
\def\Me{\check{e}}
\def\Fa{\hat{a}}
\def\Fb{\hat{b}}
\def\Fc{\hat{c}}
\def\Fd{\hat{d}}
\def\Fh{\hat{h}}
\def\Ba{\bar{a}}
\def\BB{\lambda}
\newcommand{\bra}[1]{\mbox{$\langle #1 |$}}
\newcommand{\ket}[1]{\mbox{$| #1 \rangle$}}

\begin{document}

\title{Ground state entanglement in quantum spin chains}

\author{J. I. Latorre} \affiliation{Departament d'Estructura i Constituents
de la Mat\` eria, Universitat de Barcelona, 08028, Barcelona, Spain.}
\author{E. Rico} \affiliation{Departament d'Estructura i Constituents
de la Mat\` eria, Universitat de Barcelona, 08028, Barcelona, Spain.}
\author{G. Vidal} \affiliation{Institute for Quantum Information,
California Institute for Technology, Pasadena, CA 91125 USA}
\date{\today}

\begin{abstract}

A microscopic calculation of ground state entanglement for the XY and
Heisenberg models shows the emergence of universal scaling behavior at
quantum phase transitions.  Entanglement is thus controlled by
conformal symmetry.  Away from the critical point, entanglement gets
saturated by a mass scale. Results borrowed from conformal field
theory imply irreversibility of entanglement loss along
renormalization group trajectories.  Entanglement does not saturate in
higher dimensions which appears to limit the success of the density
matrix renormalization group technique.  A possible connection between
majorization and renormalization group irreversibility emerges from
our numerical analysis.

\end{abstract}

\pacs{03.67.-a, 03.65.Ud, 03.67.Hk}

\maketitle

\section{Introduction}


At zero temperature, the properties of a quantum many-body system are dictated by the structure of its ground state. The degree of complexity of this structure varies from system to system. It ranges from exceptionally simple cases --e.g. when an intense magnetic field aligns all the spins of a ferromagnet along its direction, producing a product or unentangled state-- to more intricate situations where entanglement pervades the ground state of the system. 
Thus, entanglement appears naturally in low temperature quantum many-body physics, and it is at the core of relevant quantum phenomena, such as superconductivity~\cite{BCS57}, quantum Hall effect~\cite{La83} and quantum phase transitions \cite{sach99}.


There are several good reasons to study entanglement in quantum
many-body systems.
On the one hand, over the last decade entanglement has been realized
to be a crucial resource to process and send information in novel ways
\cite{BeDi00}. It is, for instance, the key ingredient in quantum
information tasks such as quantum teleportation and superdense coding,
and it also appears in most proposed algorithms for quantum
computation \cite{book}. This has triggered substantial experimental
efforts to produce entanglement in engineered quantum systems
\cite{FortPhys}. Consequently, one may want to investigate and characterize entanglement in those systems where it appears in a natural way, with
a view either to extract it to process quantum information or else to gain insight into physical mechanisms that can be used to entangle a
large number of quantum systems.

But one can also motivate these studies without referring to potential
applications of entanglement as a resource in quantum information
processing. The ground state of a typical quantum many-body system
consists of a superposition of a huge number of product
states. Understanding this structure is equivalent to establishing how subsystems are interrelated, which in turn is what
determines many of the relevant properties of the system. In this
sense, the study of multipartite entanglement offers an attractive
theoretical framework from which one may be able to go beyond
customary approaches to the physics of quantum collective phenomena \cite{pres01}. Most promisingly, a theory of entanglement in quantum many-body systems may also lead to the development of new numerical techniques. In particular, recent results \cite{gvprep} show how to efficiently simulate certain quantum systems through a suitable parametrization of quantum superpositions.


In this paper we present a quantitative analysis of entanglement in
several one-dimensional spin models, expanding and complementing the
results of Ref. \cite{vid02}.
The models we discuss fulfill a convenient combination of
requirements: they are solvable --the ground state can be computed by
using well-known analytical and numerical techniques-- and
at the same time they successfully describe a rich spectrum of
physical phenomena, which include ordered and disordered magnetic
phases connected by a quantum phase transition \cite{sach99}.
The paper has been divided into five more sections. A brief summary of
them follows.

The study of entanglement in a system with many particles can be
approached in several complementary ways
\cite{vid02,NiTe,W00,OW01,ved01,ved02,WFS01,WaZa02,Wa02,rest,OsNi02,Os02}. In section \ref{sec:measure} we briefly review some of them and explain the specific aspect on which we focus here. We consider a quantum spin chain in its ground state. Our aim is to determine the degree of entanglement between a {\em block} of spins and the rest of the chain, as measured by the von Neumann
entropy of the block, and to investigate how this entanglement grows with the size of the block. 


Sections \ref{sec:xy} and \ref{sec:xxz} are devoted, respectively, to
computing the entropy of a spin block for the XY model and for the XXZ
model. The calculation is divided into two parts. First, we need to construct
the ground state of the spin chain, which for general chains is a
highly non-trivial problem. Fortunately, the XY model can be treated
analytically in the limit of an infinite chain \cite{Ann,Kat,Bar}.
Similarly, for the XXZ model there are known techniques \cite{bet01}) to
easily cope with chains consisting of up to twenty spins.
Then, from the ground state of each model we extract the entropy of a
spin block. For the XY model this is shown to build down to
diagonalizing a matrix whose dimensions grow only quadratically with
the size of the block. In this way we compute the entanglement for
blocks of up to several hundreds of spins. In the XXZ model, instead, we
only consider blocks of up to ten qubits, but the results can already
be convincingly interpreted as an independent confirmation of the
conclusions drawn from the XY model.

In section \ref{sec:scaling} we discuss the findings of the previous
two sections, that can be summarized as follows: 

\begin{itemize}

\item Off a critical point, the entanglement of a block of spins with the chain --a function that turns out to grow monotonically with the number of spins in the block-- achieves a finite {\em saturation} value for sufficiently large blocks.

\item At a {\em quantum phase transition}, instead, the entropy of a block of spins grows unboundedly. More
specifically, the entropy for a critical chain grows logarithmically
in the size of the spin block, with a multiplicative coefficient that
depends only on the universality class of the phase transition. That
is, at the critical point entanglement obeys a {\em universal scaling law}.

\end{itemize}

Interestingly, the behavior of the entanglement in a critical spin
chain matches well-known results in conformal field theory, where the
geometric entropy --analogous to the spin-block entropy, but defined
in the continuum-- has been computed for 1+1 dimensional theories
\cite{Sr93,Ca94,Fi94,HoLaWi}. The geometric entropy grows
logarithmically with the size of the interval under consideration and
with a multiplicative constant given by the {\em central charge} of
the theory. As described in section \ref{sec:cft}, a consistent
picture arises. At a critical point the large-scale behavior of a spin
chain is universal, with the quantum phase transition belonging to a given universality
class. All long-range properties of the chain are then described by
the conformal field theory associated with that universality class. In particular the entanglement between a large block of spins and the rest
of the chain follows the same law as the geometric entropy.

The above connection between entanglement and the geometric entropy of
conformal theories has several implications.  Previous
calculations of the  entropy in higher dimensions
indicate that in 2- and 3-dimensional spin lattices the entropy of a
spin block grows as the size of the boundary of the block ---the same
law holding both for critical and non-critical lattices. The lack
of saturation of the entropy as a function of the size of the block explains the failure of the DMRG technique \cite{white}. Thus, we interpret
in terms of entanglement why this numerical technique ---so successful for
non-critical spin chains--- deteriorates in
critical one-dimensional lattices and 
fails to work properly both for critical and non-critical
chains and in 2- and 3-dimensional lattices \cite{RO99}.

The scaling law obeyed by the entanglement of a critical ground state implies that the greater a spin block is, the more disordered or mixed its density matrix. Thus, the entropy indicates an ordering of the reduced density matrices, according to how mixed they are. This ordering can be further refined and shown to actually emerge from majorization relations between the reduced density matrices.

Finally, we translate the results related to the c-theorem
\cite{zam86} to
quantum information.  Entanglement is
argued to decrease along renormalization group trajectories.
A number of 
numerical and analytical results are consistent with the
idea that irreversibility of renormalization group flows may
be rooted on a majorization ordering of the 
vacuum density matrices along the flow.

\section{Entanglement measures in a quantum spin chain}

\label{sec:measure}

A major difficulty in studying the entanglement in a many-body quantum
system comes from the fact that the number of degrees of freedom involved grows
exponentially with the number of interacting subsystems. In
particular, the task of computing explicitly the ground state of a chain consisting of a large number of spins, to then analyze its entanglement
properties, turns out to be very difficult, if not insurmountable.

Fortunately, for some specific spin models the ground state has been
previously computed. One can then try to characterize entanglement in
these models. This is again a rather ambitious enterprise. On the
one hand, it entails serious computational difficulties, since the
corresponding ground states, when expressed in a local basis, still
involve exponentially many coefficients. On the other hand such
characterization is also challenging from a conceptual viewpoint. The
study of entanglement of a large number of particles is a relatively
unexplored subject and it has not yet even been established what
aspects of a ground state a sensible characterization should
consider. Therefore, an important part of the problem is to actually
identify which quantities may be of interest.

This section is devoted to describe and motivate our particular
approach, which attempts to characterize the ground state of the spin
chain through the spectral properties of the reduced density matrix
for a block of spins, and in particular through its entropy. We start by presenting some generalities and reviewing previous work.

\subsection{Overview of previous work}

A state $\ket{\Psi} \in {\H_2}^{\otimes N}$ of $N$ spins is entangled
if it cannot be written as the tensor product of single-spin states,
\be \ket{\Psi} \neq \ket{\psi_1}\otimes\ket{\psi_2} \otimes \cdots
\otimes \ket{\psi_N}.  \ee Product states depend on ${\cal O}(N)$ parameters
and are therefore just a subset of zero measure in the set of states
of $N$ spins. A generic entangled state, when expressed in a local basis, depends on ${\cal O}(2^N)$ parameters. Characterizing entanglement is about identifying a reduced subset of parameters that are particularly relevant from a physical or computational point of view.

\subsubsection{Entanglement under local manipulation}

In recent years a quantitative theory of bipartite entanglement has
been developed. As proposed in the pioneering work by Bennett, Bernstein, Popescu and Schumacher \cite{benn01}, this theory is based on the possibility of converting one entangled
state $\ket{\Psi}$ into another entangled state $\ket{\Psi'}$ by
applying local operations on each of the subsystems and communicating classically, a set of transformations denoted as LOCC (see \cite{qic} for extensive reviews). The basic idea is that if the state $\ket{\Psi}$ can be converted into the state $\ket{\Psi'}$ by LOCC, 
\be
\ket{\Psi} \longrightarrow \ket{\Psi'}
\label{eq:psipsi}
\ee
then $\ket{\Psi}$ cannot be less entangled than $\ket{\Psi'}$, since LOCC can only introduce classical correlations between the subsystems. Local convertibility can in this way be used to compare the amount of entanglement in different states.

Following these ideas two remarkably simple characterizations of pure-state entanglement are possible for systems with $N=2$ subsystems:

($i$) Bennett et al. \cite{benn01} showed that, in an asymptotic sense, any entangled state $\ket{\Psi}_{AB}$ of two particles $A$ and $B$ is equivalent (that is, reversibly convertible by LOCC) to some fraction $E(\Psi)$ of an EPR state,
\be 
\frac{1}{\sqrt{2}}(\ket{0}_A\otimes \ket{0}_B + \ket{1}_A\otimes
\ket{1}_B).  
\ee 
Here the {\em entropy of entanglement} $E(\Psi)$ corresponds to the von Neumann entropy of the reduced density matrix $\rho_A \equiv \tr_{B}\proj{\Psi}$ for any one of the systems, 
\be 
E(\Psi) \equiv -\tr(\rho_A\log_2 \rho_A).
\label{eq:entofent}
\ee 

($ii$) Nielsen \cite{Nie} subsequently showed that deterministic conversions of a single copy of $\ket{\Psi}$ into $\ket{\Psi'}$ by LOCC are ruled by the majorization relation (to be introduced later in Eq. (\ref{eq:majo})). More general LOCC transformations are similarly ruled by a finite set of {\em entanglement monotones} \cite{mono}.

The optimal local manipulation of a bipartite system in a pure state is presently well understood. Bipartite pure-state entanglement can be characterized by a single measure $E(\Psi)$ in the asymptotic regime and by a small set of entanglement monotones in the single-copy case. However, none of these results has been successfully generalized to systems with $N>2$ subsystems\footnote{
The lack of complete generalizations of bipartite results to $N>2$ subsystems has a simple explanation, at least for single-copy conversions. The most allowing scenario for local manipulation of entanglement in the single-copy regime is that of stochastic local operations SLO, where the conversion of Eq. (\ref{eq:psipsi}) is only required to succeed with some non-vanishing probability. Clearly, if a conversion is not possible by SLO, then it is also not possible by LOCC. But for systems with $N\geq 3$ subsystems (and with the exceptional case of three qubits), two randomly chosen states $\ket{\Psi}$ and $\ket{\Psi'}$ are generically unconnected by SLO \cite{Dur}. This can be understood by noticing that the total number of parameters accessible to local manipulation grows linearly with the number $N$ of subsystems (the most general SLO operation can be implemented by a single measurement on each subsystem), whereas $\ket{\Psi}$ and $\ket{\Psi'}$ depend on exponentially many parameters. 
}.
In spite of the remarkable success achieved for bipartite systems, LOCC transformations do not seem to be a good a guidance to
comprehensively characterize multipartite entanglement\footnote{ 
There are many other possibilities to be considered instead, involving a coarse-grained look at entanglement. One could
base a characterization of entanglement in quantum many-body systems on the operational complexity of preparing a quantum state $\ket{\Psi}$ (or a series of quantum states $\{\ket{\Psi_N}\}$ involving an increasing number of particles $N$) by only two-particle unitary operations. The entanglement of states $\ket{\Psi_N}$ and $\ket{\Psi'_N}$ would be comparable if, say, it only takes poly($N$) two-particle operations to interconvert them.  Parameter counting shows that one can then distinguish between the class of states that can be produced with poly($N$) two-particle operations and, for instance, those requiring exp($N$) operations. Alternatively, one can study the computational cost of classically simulating the state of a many-body quantum system and its dynamics \cite{Vid03}. A third possibility is to consider how entanglement is affected by a change of scale in the system.}. 
Nevertheless, the above results can still be used to characterize any {\em bipartite} aspects of the entanglement in a multipartite system, and therefore will be sufficing for the purposes of this paper.

\subsubsection{Entanglement in spin chains}

The study of entanglement in condensed matter systems was initiated by Nielsen \cite{NiTe}. He originally analyzed two interacting spins in the Heisenberg model with an external magnetic field and studied how entanglement depends on the temperature and the intensity of the spin-spin interaction and magnetic field. In his calculations, Nielsen used Wooters' concurrence \cite{woo01}, a measure of mixed-state entanglement defined for two-qubit systems. 

More recently, several other authors have also studied the concurrence in spin systems. Wooters \cite{W00} has studied the maximal nearest neighbor concurrence that an infinite, translationally invariant spin chain can have, a result extended by O'Connor and Wooters \cite{OW01} to finite spin rings. Furthermore, concurrence in the two-spin Heisenberg model has been reanalyzed by Arnesen, Bose and Vedral \cite{ved01}. Gunlycke, Bose, Kendon and Vedral \cite{ved02} have considered a ring of several spins with Ising interaction and external magnetic field, and studied how two-spin entanglement depended on the orientation of the magnetic field. Wang, Fu, Solomon \cite{WFS01} have studied the anisotropic Heisenberg model with three spins. For a Heisenberg ring of $N$ spins, Wang and Zanardi \cite{WaZa02} have expressed the nearest neighbor concurrence in terms of the internal energy of the ring and analyzed the violation of Bell inequalities, and Wang \cite{Wa02} has investigated the concurrence in the XX model. Ref. \cite{rest} contains some other related works.

Osterloh, Amico, Falci and Fazio \cite{Os02} and Osborne and Nielsen \cite{OsNi02} have recently studied entanglement in the ground state of an infinite XY and Ising spin chains and its relation to quantum phase transitions. More specifically, they have computed the concurrence between pairs of spins, for different choices of the pair. For the Ising model with transverse magnetic field, the concurrence at nearest and next to nearest neighbors has a maximum near the critical point. Osterloh et al. have suggestively noticed that there seems to be some form of universal scaling in the derivative of the concurrence of nearest neighboring spins, and similarly for the second derivative of the next to nearest neighbors. A notable fact is that the concurrence seems to disappear at third nearest neighbors. As pointed out by Osborne and Nielsen, this can be interpreted in terms of the monogamy of entanglement \cite{Woo03}. In practice, it can also be understood as a shortcoming for using a two-qubit measure in order to capture the global distribution of entanglement along the chain.

\subsection{Entropy of a block of spins}

The approach we follow here has been proposed by Vidal, Latorre, Rico and Kitaev \cite{vid02} and, as previous works based on the concurrence, it is focussed on bipartite entanglement. But instead of analyzing the entanglement between two of the spins of the system, we consider a whole block of adjacent spins and study its
entanglement with the rest of the chain. We are particularly
interested in how the entanglement between the block and the chain
depends on the size of the block. In this way, we expect to be able to
explore the behavior of quantum correlations at different length
scales and to capture the emergence of universal scaling at a quantum critical point. Later on in this section we will further motivate our choice by referring to the relationship between entanglement and the efficiency of numerical schemes for the simulation of spin chains.

Let $\ket{\Psi_g}$ denote the ground state of a chain of $N$ spins and
let $\rho_L$, \be \rho_L\equiv \tr_{N-L} \proj{\Psi_g}, \ee be the
reduced density matrix for $L$ contiguous spins. In the models we
shall discuss, the ground state $\ket{\Psi_g}$ is translationally
invariant, so that $\rho_L$ does not depend on the position of the
block of spins but only on its size $L$. Because the chain is in a
pure state, all the information about the entanglement between the
block of spins and the rest of the chain is contained in the
eigenvalues of $\rho_L$. If $\rho_L$ is a pure state itself, then the
block is unentangled from the chain. Instead, if $\rho_L$ has many
non-zero eigenvalues, this roughly indicates a lot of entanglement
between the block and the chain. As a matter of fact, the whole
spectrum Sp$(\rho_L)$ is of interest, as we shall discuss shortly. For
concreteness, however, we will mainly use a single function of the spectrum of $\rho_L$, namely 
 the von Neumann entropy\footnote{Using the logarithm to base 2, the entropy is measured in units of information or bits. } 
\be 
S_L\equiv - \tr \left(\rho_L \log_2{\rho_L} \right),  
\ee
as a measure of entanglement.
This choice corresponds to the {\em entropy of entanglement}~\cite{benn01} (recall Eq. (\ref{eq:entofent})) between the block and the rest of the chain.

\begin{figure}[!ht]
\resizebox{!}{4.2cm}{\includegraphics{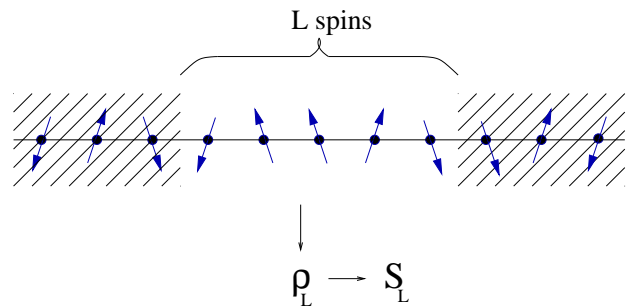}}
\caption{\label{spins} The entropy $S_L$ corresponds to the von Neumann entropy of the reduced density matrix $\rho_L$ for a block of $L$ adjacent spins, and measures the entanglement between the block and the rest of the chain. State $\rho_L$ is
obtained from the ground state $\ket{\Psi_g}$ of the N spin chain by tracing out all $N-L$ spins outside the block.}
\end{figure}

\subsubsection{Properties of the entropy of a block of spins}

Let us discuss some general properties of $S_L$ as a function of
$L$. $S_L$ is positive by construction, \be S_L \geq 0,
~~~~~ L=0,1,\cdots, N, \ee where for convenience we define $S_0\equiv
0$. Because the chain is in a pure state $\ket{\Psi_g}$, the spectrum Sp($\rho_L$) of the reduced
density matrix for a block of spins and the spectrum for the rest of the chain are the same. In particular, the two parts will also have the same
entropy. Recalling that the ground state $\ket{\Psi_g}$ is translationally invariant,
we have \be S_{L} = S_{N-L}, ~~~~~ L=0,1, \cdots, N.  \ee 
In addition, $S_L$ is a concave function \cite{petz}, \be S_L \geq
\frac{S_{L-M}+S_{L+M}}{2},
\label{eq:concavity}
\ee where $L=0, \cdots, N,$ and $M=0, \cdots,
\min~\{N\!-\!L,L\}$. This can be proved with the help of the strong
subadditivity of the von Neumann entropy~\cite{book,lie01}, 
\be
S(ABC) + S(B) \leq S(AB) + S(BC),
\label{eq:strongsub}
\ee 
where $A$, $B$ and $C$ are three subsystems and, say, $S(AB)$
denotes the entropy of $\rho_{AB}$, the joint state of systems $A$ and
$B$. Let $A$, $B$ and $C$ correspond to three adjacent blocks of our
translational invariant spin chain, with $M$, $L\!-\!M$ and $M$ spins
respectively. Then we have 
\bea S(ABC) &=& S_{L+M}\\
S(AB)=S(BC)&=&S_{L}\\ S(B)&=&S_{L-M}, \eea and
Eq. (\ref{eq:strongsub}) reads \be S_{L+M}+S_{L-M}\leq 2S_{L}, \ee
from where Eq. (\ref{eq:concavity}) follows.

Finally, we note that the above properties imply that $S_L$ does not
decrease as a function of $L$ in the interval $L \in [0,N/2]$. In
particular, in the limit of an infinite chain, $N \rightarrow \infty$,
$S_L$ becomes a non-decreasing, concave function for all finite values
of $L$.

\subsubsection{Examples}

\begin{center}
\begin{figure}[!ht]
\resizebox{!}{3.7cm}{\includegraphics{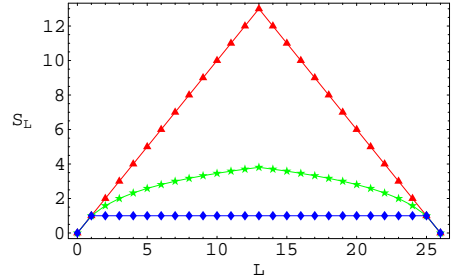}}
\caption{Bounds for the entropy $S_L$ for some pure states in a system with N=26 spins. Triangles correspond to the linear upper bound (\ref{eq:bound}), and applies to translationally invariant states (and, more generally, to arbitrary $N$-qubit states). Stars are the logarithmic upper bound for a symmetric state under permutations. The diamonds are the values of the entropy for a GHZ state.}
\label{fig:entstates}
\end{figure}
\end{center}

It is possible to get extra insight into the properties of $S_L$ as a
measure of entanglement by analyzing some particular cases, as
illustrated in Fig. (\ref{fig:entstates}).  We note first that $S_L$
is upper bounded by \be S_L \leq \min\{L, N\!-\!L\},
\label{eq:bound}
\ee since $\rho_L$ is supported in a local space of dimension
$d_L=\min\{2^{L},2^{N-L}\}$, whereas $S_L$ vanishes for all $L$ {\em
only} for product (i.e. unentangled) states.

The paradigmatic GHZ state of $N$ spins or qubits, \be
\frac{1}{\sqrt{2}}(\ket{0}^{\otimes N} + \ket{1}^{\otimes N}), \ee is
often regarded as a maximally entangled state. However, from the
present perspective it is only slightly entangled. Indeed, the
entropies of a block of spins are $S_L = 1$ for $L=1,\cdots, N\!-\!1$,
and are therefore far below the upper bound (\ref{eq:bound}).

Here we will be concerned with the ground state $\ket{\Psi_g}$ of spin chains that are
invariant under discrete translations by any number of sites. [For finite chains, we will assume that
the extremal spins are connected (spin rings) and will require invariance under
circular translations.] One could expect that translational symmetry of $\ket{\Psi_g}$ implies a more restrictive bound for the values $S_L$ can achieve. However this is not the case, since Stelmachovic et al. \cite{ste01} have found a translationally invariant state that saturates (\ref{eq:bound}). This is in contrast with the case of
states that are invariant under all possible permutations of the
spins. There the dimension $d_L^{\rm sym} = L+1$ of the symmetric
subspace leads to the upper bound $S_L^{\rm sym}\leq \log_2 L+1$.

Finally, at a critical point the ground state may have some extra
symmetries. In particular, it is known that in the large scale limit
(that is, for scales much larger than the distance between neighboring
spins) a critical spin chain is {\em conformal invariant}. We will
explore the implications of this additional symmetry in section
\ref{sec:cft}. The ground state entropy for critical chains will turn
out to grow as $S_L = k \log_2 L$ for some universal constant $k$.

\subsubsection{Majorization and von Neumann entropy}

As mentioned above, the entanglement between a block of spins and the
rest of the chain is a function of the spectrum Sp$(\rho_L)$ of the reduced density
matrix of the block. Our ultimate aim is to characterize how this
entanglement depends on the number $L$ of spins in the block. A main motivation for this is that in this way we hope to capture the emergence of universal scaling for entanglement at a quantum phase transition. Therefore we would like to be able to compare the spectrum Sp($\rho_L$) for different values of $L$.

The entropy $S_L$ can be used for this purpose, for it establishes an
order in the set of probability distributions --equivalently, in the set
of spectra of density matrices. For instance, we have mentioned above
that the entropy $S_L$ is non-decreasing in the interval
$L\in[0,N/2]$. We can now use this result to say that, according to
the entropy, the entanglement of a block of spins and the rest of the
chain monotonically increases with the size of the block (for blocks
smaller than half of the chain).

Nevertheless, there are other powerful tools to compare probability
distributions, and by using them one may obtain a finer
characterization of entanglement. In particular, a far more tight
sense of (partial) ordering between probability distributions is
established by the majorization relation \cite{Bh96}, a set of inequalities that control the conversion of bipartite entanglement by LOCC in the single-copy scenario \cite{Nie,mono,gui02}.

Let us briefly recall that a given probability distribution $x\equiv \{x_i\}$ 
(where $x_1\geq \cdots \geq x_n$) is majorized by another probability
distribution $y\equiv \{y_i\}$ (where $y_1\geq \cdots \geq y_n$), denoted $x
\prec y$, when the following series of inequalities are simultaneously fulfilled:
\bea 
x_1 &\leq& y_1 \nonumber \\ 
x_1 + x_2 &\leq& y_1 + y_2 \nonumber \\ 
&\vdots& \nonumber \\
x_1 + x_2 + \cdots + x_n &=& y_1 + y_2 + \cdots + y_n.
\label{eq:majo}
\eea 
The majorization relation $x \prec y$ expresses the fact that $y$ is {\em more
ordered} than $x$. Given two arbitrary probability distributions
$x$ and $y$, inequalities (\ref{eq:majo}) are not likely to
be simultaneously fulfilled, but when they are, most measures of order are consistent
with $x\prec y$. In particular, the von Neumann entropy fulfills \be
\rho \prec\rho'~ \Rightarrow ~ S(\rho) > S(\rho'), \ee where $\rho
\prec \rho'$ refers to majorization between the spectra of these two
density matrices.

In section \ref{sec:cft} we shall explore whether a majorization relation underlies the scaling behavior of the entropy $S_L$ at critical points. 

\subsection{Entanglement \\ in numerical studies of a quantum spin chain}
\label{sec:numerical}

There are many aspects of the ground state of a spin chain that could
be taken as a guide to characterize its entanglement. Our
choice can be motivated by the role the reduced density matrix $\rho_L$ of a
block of spins plays in some numerical schemes. We finish this section by explaining how the spectrum Sp$(\rho_L)$ of $\rho_L$ determines the efficiency of White's density matrix renormalization group (DMRG) method \cite{white} and of a recently proposed simulation scheme for the simulation of quantum spin chains \cite{gvprep}.

White's DMRG method \cite{white} is a
numerical technique that has brought an enormous progress in the study
of one-dimensional quantum systems such as quantum spin chains. It allows to compute ground state energies and
correlation functions with spectacular accuracy for non-critical spin
chains. The DMRG method, however, loses its grip (as many other methods) near a critical point and fails to work for quantum spin lattices in two or three
dimensions even away from the critical point \cite{RO99}. As recently explained by Osborne and Nielsen \cite{niel01} in the language of quantum information, the degree of performance of this method is directly related to the way it accounts for entanglement.

Let us consider a large spin chain and its ground state $\ket{\Psi_g}$. The DMRG method is based on computing properties of $\ket{\Psi_g}$ by constructing an approximation to the reduced density matrix $\rho_L$ for a block of $L$ spins, for an increasing value of $L$. This is done by retaining only the
relevant degrees of freedom of the Hilbert space associated to the block of spins. Such degrees of freedom are given by the eigenvectors of $\rho_L$ with greatest weights or eigenvalues $\{p_L^i\}$, that we assume decreasingly ordered, $p_L^i \geq p_L^{i+1}$~ $(i=1,\cdots,2^N\!\!-\!1)$.

Notice that the spectrum Sp($\rho_L$) typically contains as many as $2^N$ relevant eigenvalues, in which case the computational cost of the DMRG method explodes as the block size $L$ grows. However, not all eigenvalues have the same weight, and a good approximation to $\rho_L$ may be significantly cheaper to achieve than the exact reduced density matrix. Let $\ch_L^{\epsilon}$ denote the number of eigenvalues such that
\be
\sum_{i=1}^{\ch_L^{\epsilon}} p_L^i \geq 1 -\epsilon, ~~~~~~0 \leq \epsilon \ll 1.
\ee
That is, $\ch_L^\epsilon$ is an effective rank of $\rho_L$, resulting from ignoring all the smallest eigenvalues that sum up less than $\epsilon$. Then, if we are willing to accept a degree of accuracy $\epsilon$, the DMRG method need only retain $\ch_L^\epsilon$ eigenvectors of $\rho_L$. 

The efficiency of the DMRG depends on how small $\ch_L^\epsilon$ is. In turn, $\ch_L^{\epsilon}$ depends on how fast the eigenvalues $p_L^{i}$ decay with $i$ or, relatedly, on the entanglement between the block of spins and the rest of the chain. A very spread spectrum, roughly equivalent to a lot of entanglement, translates into a large $\ch_L^{\epsilon}$ and a large computational cost. If, instead, the eigenvalues $\{p_L^i\}$ decay very fast with $i$, implying that there is not much entanglement, then $\ch_L^{\epsilon}$ is small and so is the computational cost of the DMRG method.

Therefore, by studying the spectrum of $\rho_L$, and in particular the effective rank $\ch_L^{\epsilon}$, we may be able to assess how well the DMRG
will perform for given values of the parameters (external magnetic
field, spin-spin interaction) defining a particular spin model.

On the other hand the effective rank $\ch_L^{\epsilon}$ appears also as a decisive parameter in a recently proposed numerical scheme for the classical simulation of quantum spin chains
\cite{gvprep}. In this scheme the cost of the simulation is linear in
the number $N$ of spins in the chain and grows as a small polynomial
in $\ch^{\epsilon}$, 
\be 
\ch^{\epsilon} \equiv \max_L \ch_L^{\epsilon}, 
\ee 
that is, polynomial in the maximal effective rank achieved for blocks of adjacent spins.

Summarizing, the spectrum of $\rho_L$, through the effective rank $\ch_L^{\epsilon}$, is of direct interest for the numerical study of spin chains. The entropy $S_L$ of $\rho_L$ is related to the effective rank of $\rho_L$. Indeed, we have
\be \ch_L^{\epsilon \rightarrow 0} \geq 2^{S_L}.
\label{eq:chiL}
\ee 
In addition, numerical evidence in spin chains indicates that $2^{S_L}$ also gives a rough estimate of $\ch_L^{\epsilon}$ for small $\epsilon >0$.

In section \ref{sec:scaling} we shall discuss the results we have obtained for the entropy $S_L$, both for critical and non-critical spin chains, and analogous results for spin lattices. We will conclude that the degree of performance of the DMRG method for spin systems depends on how the entanglement between a block of spins and the rest of the system scales with the size of the block.

{\em Note added:} after completing the present work we have become
aware of a number of contributions by Peschel et al \cite{Pes} that study the spectrum of the reduced density matrix $\rho_L$ also with a view to assess the performance of the DMRG.

\section{XY model}
\label{sec:xy}

In this section we study the entanglement of an infinite XY spin
chain. We start by reviewing the main features of the XY model and by
identifying some of the critical regions in the space of parameters that define the
model. Then, we proceed to compute the ground state $\ket{\Psi_g}$ of the system, from which we obtain the reduced density matrix $\rho_L$ for $L$ contiguous spins. The
knowledge of the eigenvalues of $\rho_L$ allows us to compute 
its entropy $S_L$ and, therefore, have a quantification
of entanglement in spin chains. Further information contained
in the eigenvalues of the density matrix will be explored in section \ref{sec:cft}.

The calculations of the spectrum and ground state of the XY model that appear in this section and in the appendices review previous work in spin chains. Lieb, Schultz and Mattis \cite{Ann} solved exactly the XY model without magnetic field; Katsura \cite{Kat} computed the spectrum of the XY model
with magnetic field; Barouch and McCoy \cite{Bar} obtained the
correlation function for this model. Finally, the entropy $S_L$ was computed by Vidal, Latorre, Rico and Kitaev \cite{vid02}. Here we shall present an expanded version of this computation.

\subsection{The XY Hamiltonian}

The XY model consists of a chain of $N$ spins with nearest 
neighbor interactions and an external magnetic field, as
 given by the Hamiltonian\footnote{
Notice that the sign of the interaction can be changed by applying a 180 degree rotation along the $z$ axis (in spin space) to every second spin. Since this is a local transformation, the ground state of the original and transformed XY Hamiltonians are related by local unitary operations. Therefore both ground states are equivalent as far as entanglement properties are concerned. In this sense entanglement depends on fewer details than other properties of the chain, such as the magnetization.
}
\be
H_{XY} = -\frac{1}{2}\!\sum_l \left( \frac{1\!+\!\gamma}{2} \sigma_l^x
\sigma_{l+1}^x +
 \frac{1\!-\!\gamma}{2} \sigma_l^y \sigma_{l+1}^y + \lambda\sigma_l^z  \right).
\label{eq:XY}
\ee
Here $l$ labels the $N$ spins, $\sigma_l^{\mu}$ $(\mu=x,y,z)$ 
are the Pauli matrices,
\be
\sigma^x = \left(
\begin{array}{cc}
0 & 1 \\
1 & 0 
\end{array}
\right),~ 
\sigma^y = \left(
\begin{array}{cc}
0 & -i \\
i & 0 
\end{array}
\right),~ 
\sigma^z = \left(
\begin{array}{cc}
1 & 0 \\
0 & -1 
\end{array}
\right), 
\label{eq:sigmas1}
\ee
acting on spin $l$, with
\be
[\sigma_l^{\mu},\sigma_{m}^{\nu}] = 2i 
\delta_{lm}\sum_{\tau=x,y,z} \epsilon_{\mu\nu\tau} \sigma_l^{\tau}, 
\label{eq:sigmas2}
\ee
whereas parameter $\lambda$ is the intensity 
of the magnetic field, applied in the $z$ direction, 
and parameter $\gamma$ determines the degree of 
anisotropy of spin-spin interaction, which is 
restricted to the $xy$ plane in spin space. 

The XY model encompasses two other well-known spin models. If the
interaction is restricted to the $x$ direction in spin space, that is
$\gamma=1$, then $H_{XY}$ turns into the Ising Hamiltonian with
transverse magnetic field,
\be
H_{\rm Ising} = -\frac{1}{2}\!\sum_l
 \left( \sigma_l^x \sigma_{l+1}^x + \lambda\sigma_l^z  \right).
\label{eq:Ising}
\ee
If, instead, we consider the interaction to be
 isotropic in the $xy$ plane, $\gamma=0$, then we
 recover the XX Hamiltonian with transverse magnetic field,
\be
H_{XX} = -\frac{1}{2}\!\sum_l \left( \frac{1}{2} [\sigma_l^x 
\sigma_{l+1}^x + \sigma_l^y \sigma_{l+1}^y] + \lambda\sigma_l^z  \right). 
\label{eq:XX}
\ee

We note that these Hamiltonians are used to model
 the physics of one dimensional arrays of spins,
 but also to describe other quantum phenomena. 
For instance, the XX Hamiltonian corresponds to a 
particular limit of the boson Hubbard model
\bea
H_{B} = \sum_l \left(-w [\Ba_l^{\dagger}\Ba_{l+1} + 
\Ba_l^{\dagger}\Ba_{l+1}] - \mu n_l + \frac{U}{2} n_l(n_l-1)\right), \nonumber
\eea
where $\Ba$ are bosonic annihilation operators, 
\be
[\Ba_l,\Ba_{m}^{\dagger}]=\delta_{lm},
\ee
 and $n_l \equiv \Ba_l^{\dagger}\Ba_l$ are number operators. The boson
 Hubbard model (see chapters 10 and 11 of \cite{sach99}) consists of
 spinless bosons on $N$ sites, representing, say, Cooper pairs of
 electrons undergoing Josephson tunneling between superconducting
 islands or helium atoms moving on a substrate. The first term in
 $H_B$, proportional to $w$, allows hopping of bosons from site to
 site. The second term determines the total number of bosons in the
 model, with $\mu$ the chemical potential. The last term, with $U>0$,
 is a repulsive on-site interaction between bosons. Now, in the limit
 of large $U$, no more than 1 boson will be present at each
 site. Thus, each of the $N$ sites has an effective two-dimensional
 local space, and the identification
\bea
\sigma^{x}_l &=& \Ba_l + \Ba_l^{\dagger},\\
\sigma^{y}_l &=& -i(\Ba_l - \Ba_l^{\dagger}),\\
\sigma^{z}_l &=& 1-2\Ba_l^{\dagger}\Ba_l,
\eea
takes $H_B$ into
\be
H_B^{U\rightarrow \infty} =  -\!\sum_l \left( \frac{w}{2} 
[\sigma_l^x \sigma_{l+1}^x + \sigma_l^y \sigma_{l+1}^y] -
 \frac{\mu}{2}\sigma_l^z  \right),
\ee
which is the Hamiltonian of the XX spin chain model with 
transverse magnetic field of Eq (\ref{eq:XX}).


\subsection{Spectrum of $H_{XY}$ and critical properties}

In appendix \ref{app:app1} we show that the spectrum of
 Hamiltonian $H_{XY}$ in Eq. (\ref{eq:XY}) is given, in the
 limit of large $N$, by
\be
\label{eq:energy}
\Lambda_\phi = \sqrt{(\lambda-\cos{\phi})^2 + \gamma^2 \sin^2{\phi}},
\ee
where $\phi\in [-\pi,\pi]$ is a label in momentum space. This result was obtained by Katsura \cite{Kat}.

We can use the explicit expression (\ref{eq:energy}) of $\Lambda_{\phi}$ to
discuss the appearance of critical behavior in the XY model as a
function of parameters $(\gamma, \lambda)$.
\begin{center}
\begin{figure}[!ht]
\resizebox{!}{4.2cm}{\includegraphics{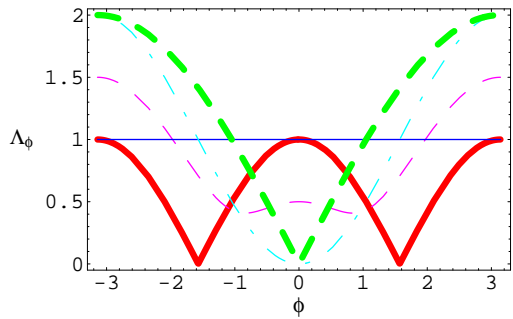}}
\caption{Energy of the system for different values of the parameters
$\lambda$ and $\gamma$ as a function of $\phi$. The thick plot
corresponds to the XX model without magnetic field, the dashed one to
a system with $\lambda=\gamma=0.5 $, the flat one is the Ising limit
without magnetic field and the dot-dashed one and the thick dashed
plot correspond to the isotropic and Ising model with $\lambda=1$,
respectively.}
\end{figure}
\end{center}
The correlation length $\xi$ characterizes the exponential 
decay of correlations in the spin chain \cite{hen01,chak01},
\be
\ev{\sigma^{a}_l\sigma^{b}_{l+L}}-\ev{\sigma^{a}_l}
\ev{\sigma^{b}_{l+L}} \sim \exp(-L/\xi), 
\ee
whereas the low energy dispersion $\Delta$ is given by
\be
\Delta \equiv \Lambda_{\phi=0}.
\ee
The critical scaling of these two quantities is characterized 
in terms of $|\lambda-\lambda_c|$ (deviation from the critical
 magnetic field $\lambda_c$) and critical exponents $\nu$ and $s$ through
\be
\begin{split}
&\xi \sim |\lambda - \lambda_c|^{-\nu}, \\ 
&\Delta \sim |\lambda - \lambda_c|^s.
\end{split}
\ee
In addition, the dynamical critical behavior is given by the energy dispersion,
\be
\Lambda_{\phi\to0} \sim \phi^z (1+(\phi \xi)^{-z}), 
\ee 
where $z$ is the dynamical exponent. An analysis of the scaling 
phenomena gives the following useful relation between the critical 
exponents: $z= s/\nu$. 

For $\lambda=1$ and any value of the anisotropy $\gamma \in [0,1]$, 
the spectrum $\Lambda_{\phi}$ in Eq. (\ref{eq:energy}) has no mass 
gap, $\Delta=0$, and the spin chain is critical, so that 
$\lambda_c=1$. As far as criticality is concerned, we need 
to distinguish two cases depending on the anisotropy  $\gamma$.

($i$) For $\gamma \in (0,1]$, Eq. (\ref{eq:energy}) implies 
that the behavior for the energy dispersions are 
\be
\begin{split}
\Delta &= \Lambda_{\phi=0} = |\lambda -1 |,
\\ \Lambda_{\phi \to0} &\sim
\sqrt{(\lambda-1)^2 + (\gamma^2+1-\lambda) \phi^2 } \\ &\sim \phi
\left(1+\frac{|\lambda-1|}{\phi} \right).
\end{split}
\ee 
At the critical point $\lambda_c=1$, we find the critical
exponents $z=1$ and $s=1$, whereas the divergence of the correlation
length in this interval is \be \xi = \frac{1}{|\lambda-1|}, \ee with a
critical exponent $\nu=1$. For later reference, we state that in 
this case the quantum spin chain belongs to the same 
{\em universality class} as the classical Ising model 
in two dimensions (or quantum Ising model in one dimension). 
The critical behavior in this class is described by the 
{\em conformal field theory} of a free massless fermion 
in 1+1 dimensions, with {\em central charge} equal to 1/2.

($ii$) For the case $\gamma=0$, we have
\be 
\Lambda_{\phi} = |\lambda-\cos{\phi}|
\ee 
and the long wave dispersion and the energy dispersion are 
\be
\begin{split}
\Delta &= \Lambda_{\phi=0} = |\lambda -1 |,
\\ \Lambda_{\phi \to0} &\sim |\lambda-1+ \phi^2/2| \\ &\sim 
\phi^2 \left|1+\frac{\lambda-1}{\phi^2} \right|.
\end{split}
\ee 
Therefore the critical point $\lambda_c=1$ leads to the
critical exponent $z=2$ and $s=1$ and the divergence of the
correlation length in this interval is 
\be 
\xi = \frac{1}{\sqrt{\lambda-1}}, 
\ee 
with a critical exponent $\nu=\1/2$. Notice, however, that in this
case the spectrum $\Lambda_\phi$ is gapless for any $\lambda\in[0,1]$,
since $\Lambda_\phi$ continuously vanishes for $\phi = \arccos
(\lambda)$. This implies that the spin chain is actually critical for
any value $\lambda\in [0,1]$ of the magnetic field. Again for later
reference, we mention that in this case the quantum spin chain belongs
to the {\em universality class} described by a free massless boson in
1+1 dimensions. This conformal theory has {\em central charge} equal
to 1.

\begin{figure}[!ht]
\resizebox{!}{6.0cm}{\includegraphics{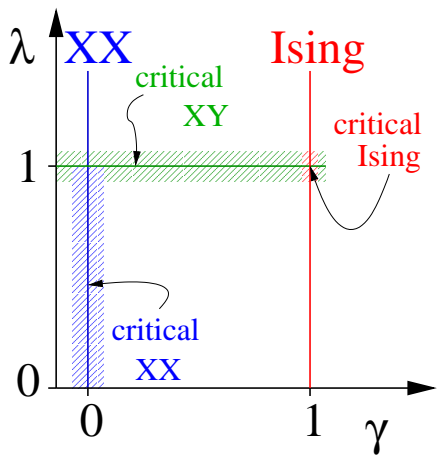}}
\caption{\label{fig:diagram} Some critical regions in the parameter space
($\gamma,\lambda$) for the XY model. The Ising model, $\gamma=1$, has
a critical point at $\lambda=1$. The XX model, $\gamma=0$, is critical
in the interval $\lambda\in [0,1]$. The whole line $\lambda=1$ is also
critical. A complete analysis of the critical regions in this model was done by Barouch and McCoy in \cite{Bar}}
\end{figure}

Summarizing, by analyzing the spectrum of $H_{XY}$ one finds two distinct critical regions in the parameter
space $(\gamma,\lambda)$, namely the line $\lambda_c =1$ and the
segment $(\gamma,\lambda)=(1,[0,1])$\footnote{
Barouch and McCoy \cite{Bar} have shown that also in the line defined by $\gamma^2+\lambda^2=1$ two-spin correlators decay as a power of the distance between spins, the signature of criticality.
}. The critical XX model
corresponds to an unstable fixed point with respect to the anisotropy
$\gamma$. If we depart from $(\gamma,\lambda) = (0,1)$ by a small
perturbation $\gamma\neq 0$, the critical behavior of the spin chain
turns from the universality class of the $XX$ model into that of the
Ising model.


\subsection{The ground state}

We now turn to determine the ground state $\ket{\Psi_g}$ of 
the XY model with open boundary conditions,
\bea
H_{XY} &=& -\frac{1}{2}\!\sum_{l=-\frac{N-1}{2}}^{\frac{N-1}{2}} \left(
 \frac{1\!+\!\gamma}{2} \sigma_l^x \sigma_{l+1}^x +
 \frac{1\!-\!\gamma}{2} \sigma_l^y \sigma_{l+1}^y \right) \nonumber\\
 &-&\frac{1}{2} \sum_{l=-\frac{N-1}{2}}^{\frac{N-1}{2}} \lambda\sigma_l^z,
\label{eq:XY2}
\eea
in the limiting case of an infinite chain, $N\rightarrow \infty$.

Through a Jordan-Wigner transformation, Hamiltonian $H_{XY}$ can be
cast into a quadratic form of fermionic operators, which in turn can
be diagonalized by means of two additional canonical transformations,
namely a Fourier transformation and a Bogoliubov transformation (see
Appendix \ref{app:app1} for details). Next we will determine the
ground state $\ket{\Psi_g}$ through a more convenient ---although
essentially equivalent--- procedure that uses Majorana operators
instead of fermionic operators. 

The present calculation was sketched in \cite{vid02} and uses the formalism described in \cite{kit}. Originally, the ground state of the XY model was determined by Lieb, Schultz and Mattis \cite{Ann} in the case of no magnetic field, and by Barouch and McCoy \cite{Bar} in the case of magnetic field.

\begin{figure} 
\resizebox{!}{6.5cm}{\includegraphics{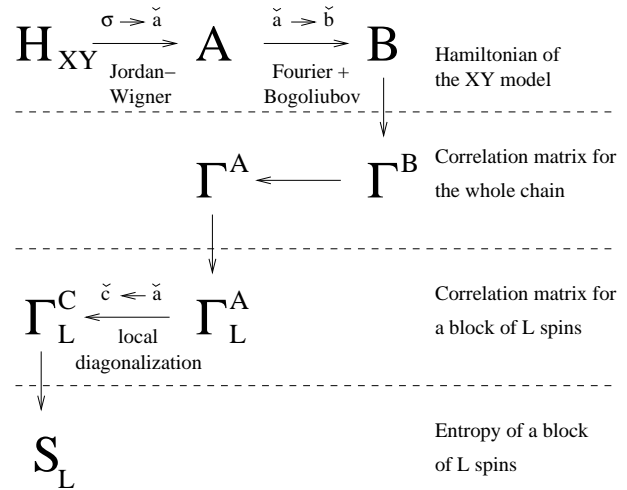}}
\caption{\label{fig:ruta} This road map describes the steps followed
in order to obtain the entropy $S_L$ of $L$ contiguous spins from an
infinite $XY$ chain. We diagonalize the Hamiltonian $H_{XY}$ by
rewriting it first in terms of Majorana operators $\Ma$ and then in
terms of Majorana operators $\Mb$. The ground state $\ket{\Psi_g}$ is
characterized by a correlation matrix $\Gamma^B$ for operators $\Mb$,
$\Gamma^A$ for operators $\Ma$. Correlation matrix $\Gamma^A_L$
describes the reduced density matrix $\rho_L$ for a block of $L$
spins. $S_L$ is finally obtained from $\Gamma_L^C$, the block-diagonal
form of $\Gamma_L^A$.  }
\end{figure}


\subsubsection{Majorana operators}

For each site $l$ of the $N$-spin chain, we consider
 two Majorana operators, $\Ma_{2l-1}$ and $\Ma_{2l}$, defined by
\be
\Ma_{2l-1} \equiv \left( \prod_{m<l} \sigma_m^z \right) \sigma_l^x; ~~~
\Ma_{2l} \equiv \left( \prod_{m<l} \sigma_m^z \right) \sigma_l^y.
\label{eq:aa}
\ee
Operators $\Ma_m$ are Hermitian and obey anti-commutation relations, 
\be
\Ma_m^{\dagger} = \Ma_m,~~~~~~\{\Ma_m, \Ma_n\}= 2\delta_{mn}.
\ee

The change of variables of Eq. (\ref{eq:aa}), parallel to the 
Jordan-Wigner transformation described in Appendix \ref{app:app1}, implies
\bea
\Ma_{2l}\Ma_{2l+1} &=& \sigma_l^y\sigma_l^z\sigma_{l+1}^x =
 i\sigma_l^x\sigma_{l+1}^x , \label{eq:firstchange}\\
\Ma_{2l-1}\Ma_{2l+2} &=& \sigma_l^x\sigma_l^z\sigma_{l+1}^y
 = -i\sigma_l^y\sigma_{l+1}^y, \\
\Ma_{2l-1}\Ma_{2l} &=& \sigma_l^x\sigma_l^y=i\sigma_l^z,
\label{eq:lastchange}
\eea
so that Hamiltonian $H_{XY}$ becomes
\bea
H_{XY} &=& \frac{i}{2}\sum_{l=-\frac{N-1}{2}}^{\frac{N-1}{2}}
 \left(~\frac{1+\gamma}{2}~\Ma_{2l}\Ma_{2l+1}
-  \frac{1-\gamma}{2}~\Ma_{2l-1}\Ma_{2l+2}~ \right) \nonumber \\
&+& \frac{i}{2}\sum_{l=-\frac{N-1}{2}}^{\frac{N-1}{2}} \lambda \Ma_{2l-1}\Ma_{2l}~,
\label{eq:majXY}
\eea
or, equivalently, 
\be
H_{XY} = \frac{i}{4} \sum_{m,n=-N}^{N-1} A_{mn} \Ma_m \Ma_n,
\label{eq:majXY2}
\ee
where $A$ is a real, skew-symmetric matrix given by
\be
A = \left[
\begin{array}{cccccc}
A_0 & A_1 & & &  \\
 -A_1^T &A_0 & A_1 & &  \\
& &\ddots & & \\
& & -A_1^T &A_0 & A_1   \\
& & &-A_1^T &A_0    \\
\end{array}
\right],
\ee
and
\be
A_0= \left[\begin{array}{cc}
0 & 2\lambda \\
-2\lambda & 0
\end{array}
\right], ~~
A_1= \left[\begin{array}{cc}
0& -(1\!-\!\gamma) \\
1\!+\!\gamma & 0
\end{array}
\right].
\ee

\begin{figure} 
\resizebox{!}{2.3cm}{\includegraphics{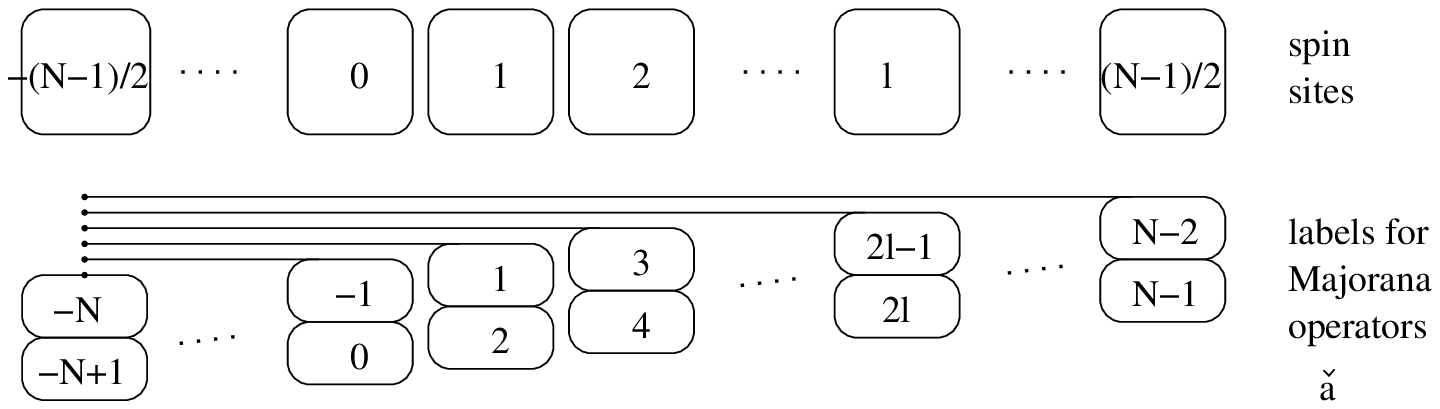}}
\caption{\label{fig:majorana} Through transformation (\ref{eq:aa}), we
can associate two Majorana operators, $\Ma_{2l-1}$ and $\Ma_{2l}$, to
site $l$ of the spin chain. Notice, however, the non-local character
of such transformation: $\Ma_{2l-1}$ and $\Ma_{2l}$ are a product of
Pauli matrices from sites $-(N-1)/2$ to $l$.  }
\end{figure}

Let $W \in SO(2N)$ be a special orthogonal matrix that brings $A$ 
into its block diagonal form $B=WAW^{T}$, 
\be
B = \bigoplus_{k=-\frac{N-1}{2}}^{\frac{N-1}{2}}\tilde{\Lambda}_k \left[
\begin{array}{cc}
0 & 1 \\
-1 &0
\end{array}
\right],
\label{eq:D}
\ee
and let 
\be
\Mb_{p} = \sum_{m=-N}^{N-1}W_{pm} \Ma_m, ~~~-N+1\leq p\leq N,
\ee
be a new set of Majorana operators, 
\be
\Mb_p^{\dagger} = \Mb_p,~~~~~~\{\Mb_p,\Mb_q\} = 2\delta_{pq}.
\ee 
The canonical transformation induced by $W$ is parallel to the 
Fourier and Bogoliubov transformations for fermionic operators
 that appear in Appendix \ref{app:app1}, where also an explicit 
expression for $\tilde{\Lambda}_k$ is displayed. In terms of 
operators $\Mb$, $H_{XY}$ reads 
\bea
H_{XY} &=&  \frac{i}{4}\sum_{p,q=-N}^{N-1} B_{pq} \Mb_p\Mb_q \\
&=& \frac{i}{4}\sum_{k=-\frac{N-1}{2}}^{\frac{N-1}{2}} \tilde{\Lambda}_k 
(\Mb_{2k-1}\Mb_{2k}-\Mb_{2k}\Mb_{2k-1}).
\eea


\subsubsection{Correlation matrix}

The diagonalization of $H_{XY}$ essentially concludes with the
determination of the explicit form of matrix $W$, as presented in
Appendix \ref{app:app2}. However, in order to analyze the resulting
ground state $\ket{\Psi_g}$, it is convenient to momentarily switch to
the more familiar language of fermionic operators. We define a set of
$N$ spinless fermionic operators $\Fb$,
\be
\Fb_k \equiv \frac{\Mb_{2k-1} + i \Mb_{2k}}{2},
\ee
$-(N-1)/2 \leq k \leq (N-1)/2$, obeying the anticommutation relations
\be
\{\Fb^{\dagger}_k,\Fb_p\}=\delta_{kp}, ~~~\{\Fb_k,\Fb_p\}=0,
\label{eq:anti_a}
\ee
in terms of which Hamiltonian $H_{XY}$ becomes, up to an irrelevant constant,
\be
H_{XY} = \sum_{k=-\frac{N-1}{2}}^{\frac{N-1}{2}} \tilde{\Lambda}_k 
\Fb_k^\dagger \Fb_k.
\ee

The ground state of $H_{XY}$ is annhilated by all $\Fb$, 
\be
\Fb_k\ket{\Psi_g}=0,
\label{eq:anihi}
\ee
so that $\bra{\Psi_g}\Fb_k^{\dagger}\Fb_k\ket{\Psi_g}$ ---that is, the
expectation value of a positive operator--- vanishes and
$\bra{\Psi_g}H_{XY}\ket{\Psi_g}=0$ corresponds to the smallest
eigenvalue of $H_{XY}$. Then, since $\Fb_k^{\dagger}\Fb_k +
\Fb_k\Fb_k^{\dagger}=I$ and $\Fb_k^{\dagger}\Fb_k\ket{\Psi_g}=0$, we
also have
\be
\Fb_k\Fb_k^{\dagger}\ket{\Psi_g} = \ket{\Psi_g}.
\ee

Let $\ev{M}$ denote the expectation value $\bra{\Psi_g}M\ket{\Psi_g}$
 for
 operator $M$. We readily have
\bea
\ev{\Fb_k} &=& 0, \label{eq:ev1}\\
\ev{\Fb_k\Fb_p} &=& 0, \\
\ev{\Fb_k\Fb_p^\dagger} &=& \delta_{kp}. \label{eq:ev3}
\eea
More generally, Wick's theorem establishes that any non-vanishing
expectation value corresponding to a product of operators $\Fb$ and
$\Fb^{\dagger}$ can be expressed in terms of
$\ev{\Fb_k\Fb_p^{\dagger}}$ and $\ev{\Fb_k\Fb_p}$ and their complex
conjugates. For instance, we have
\bea
\ev{\Fb_{k_1}\Fb_{k_2}\Fb^{\dagger}_{k_3}\Fb^{\dagger}_{k_4}}
 &=& \ev{\Fb_{k_1}\Fb_{k_2}}\ev{\Fb^{\dagger}_{k_3}\Fb^{\dagger}_{k_4}}  
- \ev{\Fb_{k_1}\Fb^{\dagger}_{k_3}}\ev{\Fb_{k_2}\Fb^{\dagger}_{k_4}}
 \nonumber \\
&+& \ev{\Fb_{k_1}\Fb^{\dagger}_{k_4}}\ev{\Fb_{k_2}\Fb^{\dagger}_{k_3}}.
\eea
This means that $\ket{\Psi_g}$ is a {\em gaussian state}, completely
characterized by the expectation values of the first and second
moments, Eqs. (\ref{eq:ev1})-(\ref{eq:ev3}).

We can now return to the Majorana operators $\Mb$. An equivalent
characterization of $\ket{\Psi_g}$ is given in terms of the
correlation matrix $\ev{\Mb_p\Mb_q} = \delta_{pq}+ i\Gamma^{B}_{pq}$,
where
\be
\Gamma^B = \bigoplus_{k=-\frac{N-1}{2}}^{\frac{N-1}{2}} \left[
\begin{array}{rc}
0 & 1  \\
- 1 &0
\end{array}
\right].
\ee
As direct substitution shows, $\Gamma^{B}$ 
amounts for both the expectation values 
$\ev{\Fb_k\Fb_p^{\dagger}}$ and $\ev{\Fb_k\Fb_p}$ 
simultaneously, which is the ultimate reason to 
conduct the present derivation in terms of Majorana operators.

Finally, we use $\Gamma^B$ to obtain the correlation 
matrix $\ev{\Ma_m\Ma_n} = \delta_{m,n} + i \Gamma^A_{mn}$ 
of the original Majorana operators $\Ma$, where 
$\Gamma^A = W^T\Gamma^BW$. As shown in Appendix \ref{app:app2}, one obtains
\be
\Gamma^A = \left[
 \begin{array}{ccccc}
\Pi_0  & \Pi_1   &   \cdots & \Pi_{N-1}  \\
-\Pi_1 & \Pi_0   & &\vdots\\

\vdots&  & \ddots&\vdots  \\
-\Pi_{N-1}& \cdots  & \cdots  & \Pi_0 
\end{array}
\right], ~~~ \Pi_l = \left[\begin{array}{cc}
0 & g_l \\
-g_{-l} & 0
\end{array}
\right],
\label{eq:GammaA}
\ee
with real coefficients $g_l$ as given, in the limit of an infinite 
chain, $N\rightarrow \infty$, by
\be
g_l = \frac{1}{2\pi}\int_{0}^{2\pi} d\phi e^{-il\phi}
\frac{ \cos \phi - \lambda - i \gamma \sin \phi}{| \cos 
\phi - \lambda - i \gamma \sin \phi|}.
\label{eq:g}
\ee

We conclude that Eqs. (\ref{eq:GammaA})-(\ref{eq:g}) contain 
a complete characterization of the ground state $\ket{\Psi_g}$ of $H_{XY}$.


\subsection{Entropy of a block of spins}

The entropy of the reduced density matrix $\rho$ for $L$ adjacent spins,
\be
S_L = -\tr (\rho\log_2\rho),
\ee
can be computed from $\Gamma^A$, Eq. (\ref{eq:GammaA}), as follows. 

In the limit of an infinite chain, the middle of the chain is fully
translational invariant, in that the same $\rho_L$ describes the state
of any block of $L$ contiguous spins. For notational convenience we
choose the block to contain qubits $l=1,\cdots,L$. We can expand the
density matrix $\rho_L$ of the block as
\be
\rho_L = 2^{-L}\!\!\!\!\!\!\!\!\!\!\sum_{\mu_1, \cdots,\mu_L =
  0,x,y,z} 
\!\!\!\rho_{\mu_1 \cdots \mu_L} ~~\sigma^{\mu_1}_{1} \cdots\sigma^{\mu_L}_{L},
\ee
where coefficients $\rho_{\mu_1 \cdots \mu_L}$ are given by
\be
\rho_{\mu_1 \cdots \mu_L} = \ev{\sigma^{\mu_1}_{1} \cdots\sigma^{\mu_L}_{L}}.
\ee

In spite of the non-local character of transformation (\ref{eq:aa}), 
the density matrix $\rho_L$ can be reconstructed from the restricted 
$2L\times 2L$ correlation matrix
\be
\ev{\Ma_m\Ma_n} = \delta_{mn} + i(\Gamma^A_L)_{mn},~~~~~m,n =1,\cdots,2L,
\ee
where
\be
\Gamma^A_L = \left[
 \begin{array}{cccc}
\Pi_0  & \Pi_1   & \cdots & \Pi_{L\!-\!1}   \\
-\Pi_1 & \Pi_0   &  & \vdots\\
\vdots & & \ddots & \vdots\\
 -\Pi_{L\!-\!1}   & \cdots &\cdots & \Pi_0 
\end{array}
\right].
\label{eq:GammaAL}
\ee
Indeed, the symmetry  
\be
\left(\prod_l \sigma_l^z \right) H_{XY} \left(\prod_l \sigma_l^z \right) = H_{XY}
\ee
implies that $\rho_{\mu_1 \cdots \mu_L}=0$ whenever the sum of $\mu$'s
equal to $x$ and of $\mu$'s equal to $y$ is odd. For instance, for
$L=4$, terms such as $\rho_{0x0z}$ and $\rho_{xy0y}$ vanish. Therefore
non-vanishing coefficients $\rho_{\mu_1 \cdots \mu_L}$ correspond to
the expectation value of a product of Pauli matrices with an even
total number of $\sigma_x$'s and $\sigma_y$'s. Such products are
mapped through the inverse of transformation (\ref{eq:aa}) into a
product of an even number of Majorana operators $\Ma_m$, with $m \in
[1,2L]$. (See Eqs. (\ref{eq:firstchange})-(\ref{eq:lastchange}) for an
example). We can then use Wick's theorem to express such products in
terms of the second moments $\ev{\Ma_m\Ma_n}$, $m,n \in [1,2L]$, all
of which are contained in $\Gamma_L^A.$

In principle, then, one could use $\Gamma_L^A$, Wick's theorem and the
inverse of transformation (\ref{eq:aa}) to compute $\rho_L$, and
extract $S_L$ from its spectral decomposition. However, the spectrum
of $\rho_L$, and its entropy $S_L$, can be computed in a more direct
way from $\Gamma_L^A$.

Let $V\in SO(2L)$ be such that it brings $\Gamma^A_L$ 
into its block-diagonal form $\Gamma^C_L = V\Gamma^{A}_LV^T$,
\be
\Gamma^C_L = \bigoplus_{l=1}^L \left[
\begin{array}{cc}
0 &\nu_l  \\
-\nu_l &0
\end{array}
\right].
\label{eq:GammaC}
\ee
Matrix V defines a set of $2L$ Majorana operators
\be
\Mc_m \equiv \sum_{n=1}^{2L} V_{mn} \Ma_n,
\ee
with correlation matrix $\ev{\Mc_m\Mc_n}$ given by 
\be
\ev{\Mc_m\Mc_n} = \delta_{mn}+i(\Gamma^C_L)_{mn}.
\ee
The structure of $\Gamma_L^C$ implies that mode $\Mc_{2l-1}$ is 
only correlated to mode $\Mc_{2l}$, a most convenient fact that 
we next exploit.

Again for the sake of clarity, we complete the present reasoning 
using a more familiar language of fermionic modes. We define 
$L$ spinless fermionic operators 
\bea
\Fc_l &\equiv& \frac{\Mc_{2l-1}+i\Mc_{2l}}{2},\\
\{\Fc_l,\Fc_m\}&=&0, ~~\{\Fc_l^{\dagger},\Fc_m\}=\delta_{lm}.
\eea
By construction they fulfill
\be
\ev{\Fc_m\Fc_n}=0, ~~~\ev{\Fc^{\dagger}_m\Fc_n} = \delta_{mn}\frac{1+\nu_m}{2},
\ee
which means that the L fermionic modes are uncorrelated, that 
is in a {\em product} state,
\be
\rho_L=\varrho_1 \otimes \cdots \otimes \varrho_L.
\label{eq:rhoprod}
\ee
[Notice that this tensor product structure does not correspond in
general to a factorization into local Hilbert spaces for the $L$
spins, but is instead a rather non-local structure].
The density matrix $\varrho_l$ has eigenvalues
\be
\frac{1\pm\nu_l}{2}
\ee
and entropy
\be
S(\varrho_l)= H_2\left(\frac{1+\nu_l}{2}\right),
\ee
where $H_2(x)= -x\log x -(1-x)\log(1-x)$ denotes the binary entropy.
The spectrum of $\rho_L$ results now from the $L$-fold product of the
spectra of the density matrices $\varrho_l$, and the entropy of
$\rho_L$ is the sum of entropies of the $L$ uncorrelated modes,
\be
S_L = \sum_{l=1}^L H_2\left(\frac{1+\nu_l}{2}\right).
\label{eq:entronu}
\ee

Summarizing: for arbitrary values of the anisotropy $\gamma$ and
magnetic field $\lambda$, and in the thermodynamic limit corresponding
to an infinite chain ($N\rightarrow \infty$), the entropy $S_L$ of the
ground state of the $XY$ model can in practice be obtained by ($i$)
evaluating Eq. (\ref{eq:g}) numerically for $l=0,\cdots, L-1$, ($ii$)
diagonalizing $\Gamma_L^A$ in Eq. (\ref{eq:GammaAL}), so as to obtain
$\nu_m$, and ($iii$) evaluating $S_L$ using Eq. (\ref{eq:entronu}).
Appendix \ref{app:app3} contains an analytical expression of the
coefficients $g_l$ in Eq. (\ref{eq:g}) for several particular cases.

\subsubsection*{Scaling of the entropy}

\begin{center}
\begin{figure}[!ht]
\resizebox{!}{4.2cm}{\includegraphics{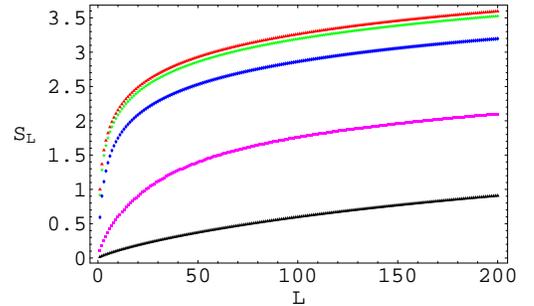}}
\caption{\label{xxmod}Entropy of the reduced density matrix for $L$
spins in the isotropic XX model, $\gamma=0$, with different external
magnetic field $\lambda$. The maximum entropy is reached when there is
no applied external field. The entropy decreases while the magnetic
field increases until $\lambda=1$ when the system reaches the
ferromagnetic limit and the ground state is a product state in the
spin basis.}
\end{figure}
\end{center}

We can now proceed to compute the entropy for the XY model with
different parameters. It is first important to note that the actual
diagonalization to be performed takes place in a $2L\times 2L$ space,
not in the huge $2^L\times 2^L$ space associated to the vacuum density
matrix. This is obvious in Fig. ~(\ref{xxmod}) where the computation can
easily include hundreds of spins.

The result obtained for the reduced density matrix of $L$ spins in the
isotropic XX model, $\gamma=0$, with no external magnetic field,
$\lambda=0$, perfectly fits a logarithmic behavior \be
S^{XX}_L=\frac{1}{3}\log_2 L + a\qquad \gamma=0,\, \lambda=0, \ee
where $a$ is a constant close to $\pi/3$. A least square fit gives an standard error of $3 \, 10^{-8}$ in the constant of the logarithmic term for this model. This result shows that
entanglement of the vacuum state scales at this critical point,
pervading the whole system and carefully organizing the complicate
superposition of states that will wind up reproducing correlators. The
scaling of entanglement, furthermore follows some universality
properties we shall discuss later.

\begin{center}
\begin{figure}[!ht]
\resizebox{!}{4.2cm}{\includegraphics{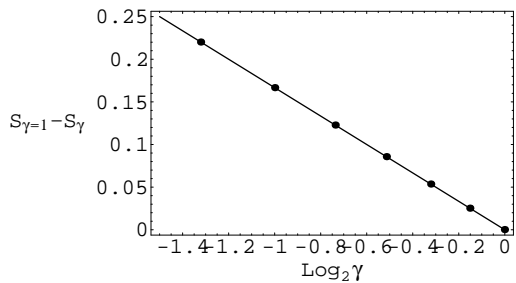}}
\caption{\label{anisot}Difference of the entropy $\lim_{L\rightarrow
\infty} \left[ S_L(\gamma=1)-S_L(\gamma) \right]$ for different values
of the anisotropy $\gamma$. For every $\gamma$ the model
is critical and the entropy scales as ${1/ 6} \log_2 L+
a(\gamma)$, where the $L$-independent function $a(\gamma)$
 is perfectly fitted by $-\frac{1}{6}
\log_2 \gamma$.}
\end{figure}
\end{center}

It is easy to extend our computation to other critical and
non-critical points in the parameter space for the XY
system. Fig.~(\ref{anisot}) shows the scaling of entanglement as we
scan $\gamma$.  Note again the logarithmic scaling of the entropy
although its coefficient is now 1/6 instead of 1/3.  The constant
correction to the logarithmic scaling is such that
\be 
S^{XY}_L=\frac{1}{6}\log_2 L+a(\gamma), 
\ee 
so that 
\be 
\lim_{L\rightarrow \infty} \left[ S_L(\gamma=1)-S_L(\gamma) \right] =
-\frac{1}{6} \log_2 \gamma \ .
\ee

\begin{center}
\begin{figure}[!ht]
\resizebox{!}{4.2cm}{\includegraphics{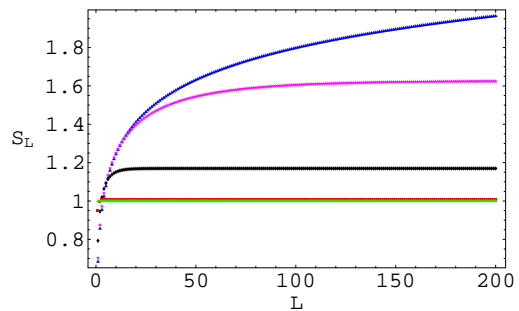}}
\caption{\label{ising}Entropy for the reduced density matrix to $L$
spins of the Ising model, $\gamma=1$, with different values of the
external magnetic field, $\lambda \in \{0,1\}$. The maximum
entropy is reached at the critical point when the applied field is
one. Other values of the magnetic field lead to saturation of the
entropy. For a magnetic field $\lambda=0$ the ground state of the
system is in the N\'eel state or GHZ state.}
\end{figure}
\end{center}

When $\gamma = 1$ the system is described by the Ising model. The
entropy behavior in this limit can be viewed in the Fig.~(\ref{ising}).
When the magnetic field is turned to $\lambda=1$ the entropy
reproduces the scaling law \be S^{Ising}_L=\frac{1}{6}\log_2 L +
a\qquad \gamma=1,\, \lambda=1.  \ee In this model, the least square fit of the logarithmic behavior gives a standard error of $4 \,10^{-9}$. For the Ising model with no
external field, the ground state that minimizes the total energy is
the N\'eel state, a macroscopic GHZ state, for which the entropy is
always equal to one.

\section{Heisenberg model}

\label{sec:xxz}

\subsection{The XXZ Hamiltonian}

The XXZ model consists of a chain of $N$ spins with nearest neighbor interactions and an external magnetic field, as given by the Hamiltonian
\be
\label{eq:XXZ}
H_{XXZ} = \sum_l \left(\frac{1}{2}[\sigma^x_l\sigma^x_{l+1} + \sigma^y_l\sigma^y_{l+1} + \Delta \sigma^z_l\sigma^z_{l+1}]  +\BB \sigma^z_l \right).
\ee
As in the previous section, $l$ labels the $N$ spins and $\sigma_l^{\mu}$ ($\mu=x,y,z$) are the Pauli matrices, Eqs. (\ref{eq:sigmas1})-(\ref{eq:sigmas2}). Parameter $\Delta$ evaluates the anisotropy, in the $z$ direction, of the antiferromagnetic Heisenberg interaction, whereas $\lambda$ is the strength of a magnetic field applied in the $z$ direction. 

The XXZ model includes as special cases two other well-known spin models. The XXX model corresponds to a fully isotropic interaction, $\Delta=1$,
\be
\label{eq:XXX}
H_{XXX} = \sum_l \left(\frac{1}{2}[\sigma^x_l\sigma^x_{l+1} + \sigma^y_l\sigma^y_{l+1} + \sigma^z_l\sigma^z_{l+1}]  +\BB \sigma^z_l \right).
\ee
Also, when the interaction in restricted to the plane $xy$ in spin space, $\Delta=0$, we recover the XX model of Eq. (\ref{eq:XX}).

These Hamiltonians are commonly used to model the physics of certain spin chains, but also to describe other quantum systems. For instance, the XXX Hamiltonian without magnetic field can be obtained in a particular limit of the fermion Hubbard model,
\be
H  = \sum_{l\tau} \left( \epsilon n_{l\tau} + t
(\Fh^\dagger_{l\tau}\Fh_{l+1\tau}+ \Fh^\dagger_{l+1\tau}
 \Fh_{l\tau}) \right) 
+ \sum_l U n_{l\uparrow} n_{l\downarrow},
\ee 
where $\Fh$ are fermionic annihilation operators,
\be
\{ \Fh_{l\tau},\Fh_{m\nu}^{\dagger}\} = \delta_{lm}\delta_{\tau\nu},
\ee
$n_{l\tau} = \Fh_{l\tau}^\dagger\Fh_{l\tau}$ are fermion number operator and $l$ labels one of $N$ sites of a chain while $\nu$ denotes one of two spin orientations, $\uparrow$ or $\downarrow$. The fermion Hubbard Hamiltonian (see chapter 10 of \cite{sach99}) was originally introduced to describe the motion of electrons in transition metals, and consists of spin-1/2 fermionic particles moving along the $N$ sites. Parameter $\epsilon$ is the energy cost of having one fermion, $t$ is the tunneling parameter and $U$ quantifies the interaction between two fermions at the same site. Notice that the total number of fermions $n=\sum_{l\tau}n_{l\tau}$ is a constant of motion. Then, when restricted to a subspace with a given value of $n$, the first term in the Hamiltonian is proportional to the identity and can be omitted. Let us consider the case when the total number of fermions is $N$, that is, the same as the number of sites. In the limit $U \gg t$, two fermions are not energetically allowed to be on the same site, and we have one fermion per site. In this limit, and with the identification
\bea
\sigma_l^x &=& \Fh_{l\uparrow}^\dagger\Fh_{l\downarrow} + \Fh_{l\uparrow}^\dagger\Fh_{l\downarrow},\\
\sigma_l^y &=& -i(\Fh_{l\uparrow}^\dagger\Fh_{l\downarrow} - \Fh_{l\uparrow}^\dagger\Fh_{l\downarrow}),\\
\sigma_l^z &=& \Fh_{l\uparrow}^\dagger\Fh_{l\uparrow} - \Fh_{l\downarrow}^\dagger\Fh_{l\downarrow},
\eea
the Hubbard model can be recast (using perturbation theory) into the isotropic Heisenberg model without magnetic field,
\be H = J \sum_l
\vec{\sigma}_l \cdot \vec{\sigma}_{l+1}, 
\ee 
where $J=4\frac{t^2}{U}$. Metal-insulator transitions, superconductive systems or magnetic properties can be explained with this Hamiltonian. 

Quantum phase transitions are identified with points of non-analyticity in the ground state energy of the spin chain. This non-analyticity may appear in the limit of a large chain or, as in the present case, may be due to level crossing. Notice that we can decompose $H_{XXZ}$, Eq. (\ref{eq:XXZ}), in terms of three commuting parts, 
\be
H_{XXZ} = H_1 + \Delta H_2 + \lambda H_3.
\ee
As we change $\Delta$ or $\lambda$, an excited state may see its energy decreased enough as to become the ground state. At one such point, the ground state energy $E(\Delta,\lambda)$ will not be analytical. Thus, in this model quantum phase transitions occurs for finite chains.

\subsection{Bethe Ansatz for the Heisenberg model}

Several properties make the Heisenberg Hamiltonian with periodic boundary conditions completely
integrable.  Two symmetries are essential to get the model solution
and are used in the \emph{Bethe Ansatz} (\emph{BA})
 \cite{bet01}. Rotational symmetry about the $z$-axis in spin space implies that the
$z$-component of the total spin $S^z_T$,
\be
S^z_T = \frac{1}{2}\sum_l \sigma_l^z,
\ee
is conserved. Sorting the basis vectors according to the quantum number $S^z_T=N/2-r$, where $N$
is the number of sites in the chain and $r$ the number of spins down,
is all that is required to block diagonalize the Hamiltonian. The second symmetry is the invariance of $H$ with
respect to discrete translations by any number of lattice spacings.
To reconstruct the whole spectrum of the Heisenberg model, Bethe's
idea is to start with the ferromagnetic state $\ket{F}$,
\be
\ket{F} = \ket{\uparrow \uparrow \cdots \uparrow},
\ee
whose spin angular momentum, $\frac{N}{2}$, is maximum, and to get a translationally invariant eigenstate of the Hamiltonian with one unit less of spin angular momentum. The rest of the spectrum of the Heisenberg model is obtained by iterating this process.

Here, we shall use the \emph{BA} to get a numerical but exact solution
of a finite spin chain. Finite size effects are present but 
can be controlled by comparing different sizes. Scaling of
entanglement is thus approached asymptotically as the size
grows. For a fixed value of $S^z_T=N/2-r$, eigenstates are of the form
\be
\label{eq1} 
|\Psi \ra = \sum_{1 \le n_1 < ... < n_r \le N}a(n_1,...,n_r)
|n_1,...,n_r\ra, 
\ee 
where $n_1,\cdots,n_r$ list the position of the $r$ spin that are down, and $a(n_1,\cdots,n_r)$ fulfills
\be a(n_1,...,n_r)=\sum_{\mathcal{P} \in
\mathcal{S}_r} \exp \left(\rm{i}\sum_{j=1}^r k_{\mathcal{P}j} n_j +
\frac{\rm{i}}{2}\sum_{i\le j} \theta_{\mathcal{P}i,\mathcal{P}j}
\right). \ee 
Here $\mathcal{P}\in \mathcal{S}_r$ denotes one of the $r!$
permutations of $\{ 1,...,r\}$ and $k_i$ and $\theta_{i,j}$ with
$(i,j)\in \{ 1,...,r\}$ are the parameters to be determined.

Three general conditions hold for these parameters \cite{orb01}:
\bea
\label{eq2}
\theta_{i,j}&=&-\theta_{j,i} ~~~~~~~~~~ \forall \{ i,j\},\nonumber \\
\cot\frac{\theta_{i,j}}{2}&=&\frac{\Delta \sin \frac{k_i-k_j}{2} }{\cos
\frac{k_i+k_j}{2}+\Delta \cos \frac{k_i-k_j}{2} } \hspace{3ex}
(i,j)\in \{ 1,...,r\}, \nonumber\\ 
N k_i &=& 2 \pi \lambda_i + \sum_{j \ne i}
\theta_{i,j} ~~~~~~~~~~ i \in \{ 1,...,r\},
\eea
where the integers $\lambda_i$ are called Bethe quantum numbers. The
states are completely determined by the Bethe quantum numbers and
Eqs. (\ref{eq1})-(\ref{eq2}).

It is known~\cite{yan01} that the set $\{ \lambda_i \}$ with the
lowest energy for each ${S}^z_T=N/2-r$ satisfies: \be \lambda_i
={S}^z_T -1 + 2i = \frac{N}{2} - r -1 + 2i \hspace{3ex} i \in \{
1,...,r\} . \ee 
 The expression for the ground state energy reads
\be E-E_F= -J \sum_{i=1}^r (\Delta - \cos k_i ) + \BB S^z_T,
\hspace{3ex} E_F= \frac{JN\Delta}{4}.  \ee

We have found convenient to solve the previous system of non-linear
equations 
using  a minimization code for absolute errors
based on genetic algorithms, although many other numerical techniques
could have been used. Precision can be controlled by
setting a maximum bound in the numerical error
 in the parameters $\theta$'s and $k$'s to
$10^{-5}$. This small error has proven enough to
detect the scaling of entropy.
The systems that we study have an even number of
sites $N$, so the eigenvalue of the $S^z_T$ operator in the
ferromagnetic system is the integer $N/2$. 

\subsection{Entropy of a block of spins}

\begin{center}
\begin{figure}[!ht]
\resizebox{!}{4.2cm}{\includegraphics{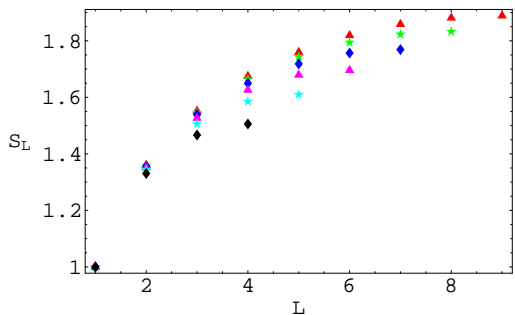}}
\caption{\label{size} Dependence of the entropy $S_L$ on the finite size of
the chain in the isotropic Heisenberg model without magnetic field. In
the plot, triangles, stars, diamonds, triangles, stars and diamonds
correspond to 18, 16, 14, 12, 10, and 8 spins chains, respectively. All curbs coincide with the upper one for low block sizes $L$, but disagree as $L$ gets closer to $N/2$ due to a finite-size saturation effect.}
\end{figure}
\end{center}

The fact that the \emph{BA} is used to compute  numerically
the eigenvalues of the reduced density matrix of blocks
of spins 
limits the size of the system that can be studied. 
The results we obtain are thus less precise than those
for the XY model although  scaling laws
can be inferred with confidence. As a first step,  
we concentrate on the effect of the finite size
system for the entropy results. More precisely, we analyze the isotropic model
without magnetic field in a chain of $N = \{8,10,12,14,16,18\}$
sites. The results are plotted in 
Fig. (\ref{size}). Finite size effects bend down the entropy 
when the size of the block  approaches half of the chain. 
Smaller blocks are less sensitive to the finite size and
show good scaling. The numerical results indicate that the
entropy behavior converges to a logarithmic scaling when 
as the size of the system increases. This asymptotic
behavior corresponds to
\be
 S_L\sim \frac{1}{3} \log_2 L\ .
\ee
As we shall discuss in the next two sections, this entropy scaling falls into the universality class of a free boson.

\subsubsection{Isotropic model in a magnetic field}

We first analyze the XXX model with a magnetic field, Eq. (\ref{eq:XXX}). We can now use the \emph{BA} to compute the eigenstates of the Hamiltonian
in the critical interval: $\lambda \in [0 , 2)$ ~\cite{bon01}.
 For values of $|\lambda|$ greater or equal to 2,
every spin in the system is pointing in the magnetic field direction
and the state is a product of $S^z_i$ eigenstates. Entanglement
has disappeared.

As implicit in the  \emph{BA}, the ground state has a well defined
magnetization, $S^z_T$. This
quantity changes every time the tuning of the magnetic field
implies a new level-crossing. The relation between the
magnetization, $S^z_T$, and the ratio $\lambda$ between the magnetic
field $\BB$ and the coupling constant $J$ is well defined and can be
easily obtained from the spectrum of the model.
\begin{center}
\begin{figure}[!ht]
\resizebox{!}{4.2cm}{\includegraphics{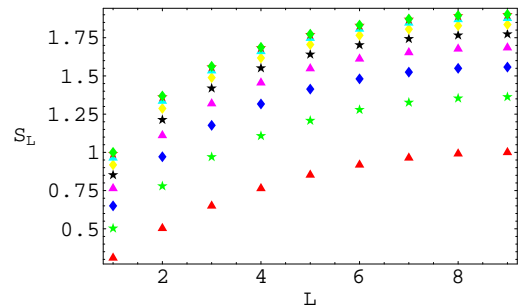}}
\caption{\label{mag18} Entropy $S_L$ in a ring of $N=18$ spins, for blocks of size up to $L=N/2=9$ spins. The curbs correspond to the values of the magnetic field, $\lambda \in \{0,
0.24, 0.68, 1.05, 1.35, 1.59, 1.77, 1.89, 1.97 \}$, for which there is level crossing. For a given $L$, the entropy remains constant in the interval between level crossing, but every time the coupling is at one of the latter points the entropy value changes. The maximum in the entropy is obtained for the
antiferromagnetic system without magnetic field, while it goes to zero as $\lambda$ approaches 2.}
\end{figure}
\end{center}

\begin{center}
\begin{figure}[!ht]
\resizebox{!}{4.2cm}{\includegraphics{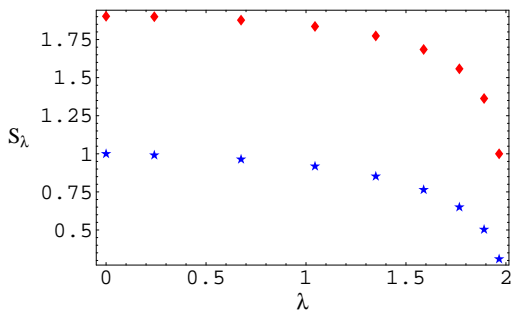}}
\caption{\label{logmag} Entropy $S_L$ in a ring of $N=18$ spins, for different
values of the magnetic field $\lambda$. Stars correspond to the reduced density matrix of one spin, $S_{L=1}$, and diamonds to the reduced density matrix for half ring, $S_{L=N / 2 = 9}$.}
\end{figure}
\end{center}

The Heisenberg model has two limiting behaviors.  On the one hand, for
$\lambda = 0$ the ground state is  antiferromagnetic,
 with a null  angular momentum eigenvalue. On the other
hand, for $\lambda \ge 2$, the ground state of the system corresponds
to the ferromagnetic state $\ket{F}$. Figs. (\ref{mag18}) and 
(\ref{logmag}) illustrate how the entropy of a
finite spin ring changes when $\lambda$ varies in the interval
$[0,2]$. Fig. (\ref{mag18})  shows the way the entropy
decreases as $\lambda$ increases, obtaining the zero value when
$\lambda \to 2$ where the ground state turns into
 a product state. The plotted values of $\lambda$ are those for which the ground state has a
level crossing for a ring with $N=18$ sites. The ground state, and therefore also its entropy, remain constant in the
interval between level crossings.
Fig. (\ref{logmag})
plots the way the entropy increases logarithmically when
$\lambda$ decreases, keeping the number of traced spins fixed.

\subsubsection{Anisotropic model}

Let us now analyze the behavior of entanglement
as a function of the anisotropy $\Delta$ in a Heisenberg chain without magnetic field, that is $H_{XXZ}$ with $\lambda=0$ in Eq. (\ref{eq:XXZ}).

In this case, the critical interval~\cite{aff01} corresponds to
 $\Delta \in [-1,1]$. For $\Delta=1$ we recover the isotropic antiferromagnetic
model, $H_{XXX}$. In the case $\Delta=-1$ the Hamiltonian can be transformed into the isotropic ferromagnetic Hamiltonian,
\be
H = -\sum_l  \vec{\sigma}_l\cdot\vec{\sigma}_{l+1},
\ee
by further rotating each second spin by 180 degrees in the z direction. Notice that entanglement remains invariant under this transformation.
This argument shows that entanglement is a robust magnitude and
depends on fewer details of the system than e.g. its
magnetization. 

\begin{center}
\begin{figure}[!ht]
\resizebox{!}{4.2cm}{\includegraphics{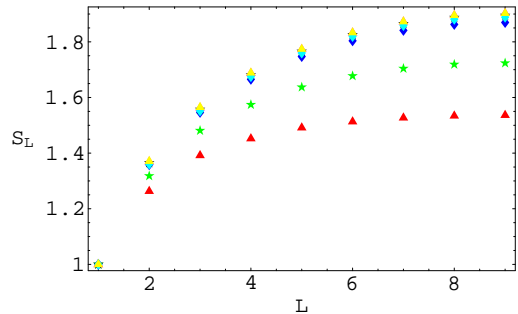}}
\caption{\label{ani18} Entropy $S_L$ in a ring of $N=18$ spins and for different values of the anisotropy $\Delta$. The critical interval is for $\Delta \in
[-1,1]$, and leads to the upper, superposed curbs. The points out of the critical interval are diamonds with
$\Delta = 1.5$, stars with $\Delta = 2.0$ and triangles with $\Delta =
2.5$. Thus, as we change the anisotropy $\Delta$ away from the critical
interval, the entropy gets saturated by a value that decreases with $\Delta$.}
\end{figure}
\end{center}

The anisotropy is a marginal deformation in the interval $\Delta \in [-1,1]$, in that for any such value of $\Delta$ the system has the same large scale behavior. Instead, for other values of $\Delta$ a gap appears in the spectrum, giving raise to a new length scale. A finite correlation length in the system takes over and all correlations decay exponentially. Fig. (\ref{ani18}) shows that in this case the entropy $S_L$ gets saturated as a function of $L$.

\section{Critical versus non-critical entanglement}

\label{sec:scaling}

We now turn to analyze the results of the calculations described
in the previous two sections.

The big picture emerging from these calculations is that there is a
clear distinction between the entanglement in a non-critical chain and
that in a critical chain, as measured by the entropy $S_L$ of the
reduced density matrix $\rho_L$ for $L$ contiguous spins, see Figs. (\ref{xxmod}) and (\ref{ising}). For all
non-critical models, $S_L$ reaches a saturation value $S^*$ as $L$ increases,
whereas for critical chains $S_L$ grows unboundedly with $L$.

\subsection{Non-critical spin chains}

Recall that in the non-critical regime, a spin chain is characterized
by a gap between the energy of the ground state and that of the first
excited state. Relatedly, correlations between increasingly distant
spins decay exponentially, 
\be
\ev{\sigma^{a}_l\sigma^{b}_{l+L}}-\ev{\sigma^{a}_l}
\ev{\sigma^{b}_{l+L}} \sim \exp(-L/\xi), 
\ee
where $\xi$ is the correlation length
 (we take the distance between nearest spins
 as unit of distance). For non-critical chains, we find that the entropy
$S_L$, a growing function with the block size $L$, is upper-bounded by
a saturation value $S^*$. This value depends on the parameters
specifying the spin chain, and becomes larger as the chain gets closer to
a critical point or critical phase. For any fixed value of the
parameters specifying the spin model, $S_L$ approaches
the saturation value $S^{*}$ for block sizes $L$ of the order of the correlation
length $\xi$. We can conclude, therefore, that the entanglement between a
block and the rest of an infinite spin chain has a fixed value for
blocks larger than the correlation length $\xi$.

Recall from Eq. (\ref{eq:bound}) that $S_L$ could in principle grow as much as $L$. This means that, for large block size $L$, the entanglement of a block is negligible when compared to its maximal possible value,
\be
\lim_{L\rightarrow \infty} \frac{S^*}{L} = 0.
\ee
Thus, the ground state $\ket{\Psi_g}$ entangles all the spins in the chain, either directly or through intermediate spins, but the system does only contain a very small amount of entanglement at any scale.

The above calculations can be used not only to obtain $S_L$, but also
to determine the whole spectrum Sp$(\rho_L)$ of decreasingly ordered eigenvalues $\{p_L^i\}$ of the reduced density matrix $\rho_L$ for the block (see section \ref{subsec:majo_and_ent}). 
For any $L$, this spectrum contains only a very
small number of relevant eigenvalues, together with many small, rapidly decaying eigenvalues with an insignificant overall weight. Thus, in spite of a
exponential growth of the rank of $\rho_L$ with $L$, the effective rank
$\ch_L^{\rm \epsilon}$ (recall section \ref{sec:numerical}) also gets saturated, with the saturation value $\ch^{\rm \epsilon}$ being reached when $L$ is of the order $\xi$.

From the point of view of numerical calculations, the saturation of $\ch_L^{\epsilon}$ can be used to interpret the extraordinary success of the DMRG
in non-critical spin chains and other one-dimensional systems. Only a
fixed, small number of eigenvectors of $\rho_L$ must be
retained in order to capture the relevant degrees of freedom as $L$
increases \cite{RO99}. Similarly, a bounded effective rank $\ch_L^{\epsilon}$ as a function of $L$ indicates that non-critical spin chains can
be efficiently simulated using the techniques introduced in \cite{gvprep}.

\subsection{Critical chains}

The energy spectrum of a critical spin chain is gapless and the
two-spin correlation function is characterized by a power scaling law,
\be
\ev{\sigma^{a}_l\sigma^{b}_{l+L}}-\ev{\sigma^{a}_l}
\ev{\sigma^{b}_{l+L}} \sim L^{-q}, 
\ee
where $q>0$ (there are cases where marginal 
deformations lead to logarithmic corrections to the power
law).  This scaling law implies an infinite correlation length
$\xi$. For critical spin chains we have obtained that $S_L$ grows
unboundedly as a function of $L$. In particular, in all the cases we
have analyzed the growth of $S_L$ is asymptotically given by \be S_L
\sim k\log_2 L,
\label{eq:tomalog}
\ee where $k$ is either $1/3$ or $1/6$ depending on the parameters of
the critical chain. Thus, for critical spin chains the entanglement
between a block of spins and the rest of a chain grows unboundedly
with the size $L$ of the block, in sharp contrast with the
non-critical case. In the next section we will elaborate on the
particular shape (\ref{eq:tomalog}) of $S_L$ for critical chains.

Given fixed computer memory and execution time, the DMRG works with
significantly smaller accuracy for critical spin chains than for
non-critical ones. This result can now be qualitatively interpreted in
terms of the arbitrarily large entanglement that links a spin block to
the rest of the chain as $L$ grows. Indeed, a larger value of $S_L$
implies that more eigenstates of $\rho_L$ must be retained in order
for the DMRG to achieve a similar accuracy as when $S_L$ is small
(recall Eq. (\ref{eq:chiL})).

We notice that even if $S_L$ for critical chains grows unboundedly with $L$, we again have that its value is much smaller than the upper bound (\ref{eq:bound}), $S_L/L$ becoming negligible for large $L$. Therefore, even if entanglement is the responsible for the appearance of long-range correlations at a critical point, we find that the amount of entanglement in the ground state of a spin chain is surprisingly small compared to its maximal possible value.

Finally, in section \ref{sec:cft} it will be argued that in $d\geq 2$
dimensions, the entanglement between a $d$-dimensional block of $L^d$ spins and the rest of a critical
spin lattice grows as $L^{d-1}$. The same scaling is expected to apply to non-critical spin lattices. This implies that
$\ch_{L}^{\rm \epsilon}$ grows exponentially with the size of the block,
explaining the systematic breakdown of the DMRG in quantum spin
systems in more than one dimension.


\section{Scaling of entanglement and conformal field theory}

\label{sec:cft}

The set of results found in the previous sections exemplify the connection
between entanglement and quantum field theory concepts. The bridge
between quantum information and quantum field theory can be further
explored and exploited in both directions.
In this sense, we shall translate
results from black hole entropy and effective actions
on gravitational backgrounds  to the language of quantum information. 
Working in the opposite direction,  quantum information
natural measures of order based on majorization theory seem to find
their way into renormalization group irreversibility. 
The cross-fertilization of these ideas requires a change of language
and quite a lot of background on quantum field theory
that we cannot cover here
in a self-contained or satisfactory way. Nevertheless, we
will very briefly review some concepts in conformal field 
theory and c-theorem necessary to state their 
spinoff in quantum information theory.


\subsection{Entropy and conformal field theory}

The study of entropy in systems with an infinite number of degrees of
freedom has received quite some attention in the context of quantum
field theory and black hole physics.  Historically, the thermodynamics
of a black hole lead to the Bekenstein-Hawking entropy (see
for instance \cite{Be94}. This entropy
is associated to the counting of microscopic degrees of
freedom inside the horizon and scales as the area of the black hole.
Some authors suggested that the origin of this entropy might be rooted
in the loss of information forced upon an external observer by the
existence of a horizon.  Although this point of view is no longer
pursued, its analysis comes naturally due to the combination of
quantum mechanics and general relativity which underlies the
holographic principle. The field theoretical definition of entropy faces
the traditional problem of renormalization in quantum field theory and
receives the name of fine-grained entropy as well as geometric
entropy.

Three computations of genuine quantum field theory entropy are worth
recalling. First, Srednicki~\cite{Sr93} considered a properly
regularized massless bosonic field theory in a universe which is
divided by an imaginary sphere of radius $R$ into its inside and
outside parts. He then numerically constructed the density matrix of
the reduced outside system and found that its entropy scaled with an
area law in 3+1 dimensions. For 1+1 dimensions, the scaling behavior
of the entropy was found to be logarithmic.  Second, Callan and
Wilczek~\cite{Ca94} put forward the concept of geometric entropy in
1+1 dimensions.  There, the power of conformal symmetry was used at
full steam in order to compute the entropy of a conformal field theory
when reduced to a finite geometry. We shall come back to this result
shortly.  Finally, the third relevant computation was carried out by
Fiola, Preskill, Strominger and Trivedi~\cite{Fi94} who mapped a
regularized field theory in 1+1 dimension to the Rindler coordinates
and recovered the logarithmic behavior of the microscopic entropy. All
these computations needed an explicit ultraviolet regulator since quantum field theory contains
infinitely many degrees of freedom. Note that in our case spin chains
come equipped with the intrinsic ultraviolet cutoff of the lattice
spacing.

Although more restrictive in scope, the computation of the geometric
entropy of a 1+1 dimensional field theory brings the advantage that
the result is casted in terms of the parameters that classify
conformal field theories. More concretely, the result found by
Holzhey, Larsen and Wilczek~\cite{HoLaWi} reads
 \be S_L=
\frac{c+\bar{c}}{6} \log_2 L \ee 
where $c$ and $\bar c$ are the so
called central charges for the holomorphic and antiholomorphic sectors
of the conformal field theory.

Let us briefly recall that conformal field theories~\cite{cft} are
classified by the representations of the conformal group in 1+1
dimensions. The operators of the theory fall into a structure of
highest weight operators and its descendants. Each highest weight
operator carries some specific scaling dimensions which dictates those
of its descendants. The operators close an algebra implemented into
the operator product expansion.  One operator is particularly
important: the energy-momentum tensor $T_{\mu\nu}$. It is convenient
to introduce holomorphic and antiholomorphic indices defined by the
combinations $T=T_{zz}$ and $\bar T=T_{\bar z \bar z}$ where $z=x^0+i
x^1$ and $\bar z=x^0-i x^1$. Denoting by $\ket{0}$ the vacuum state, the central charge of a conformal field
theory is associated to the coefficient of the correlator \be \langle
0\vert T(z) T(0)\vert 0\rangle = \frac{c}{2 z^4} \ee and the analogous
result for $\bar c$ in terms of the correlator $\langle 0\vert \bar
T(z) \bar T(0)\vert 0\rangle$.  A conformal field theory is
characterized by its central charge, the scaling dimensions and
the coefficients of the operator product expansion. Furthermore,
unitary theories with $c<1$ only exist for discrete values of $c$ and
are called minimal models. The lowest lying theory corresponds to
$c=\frac{1}{2}$ and represents the universality class of a free
fermion.

The central charge plays many roles in a conformal field theory.  It
was introduced above as the coefficient of a correlator of
energy-momentum tensors, which means that it is an observable. The
central charge also characterizes the response of a theory to a
modification of the background metric
where it is defined.  Specifically, the scale anomaly
associated to the lack of scale invariance produced by a non-trivial
background metric is 
\be \langle 0\vert T^\mu_\mu\vert 0 \ra =
-\frac{c}{12} R \ee 
where $R$ is the curvature of the background
metric. This anomaly can also be seen as the emergence of a non-local
effective action when the field theory modes are integrated in a
curved background.

The results we have found in our analysis match perfectly the
geometric entropy computation. In the case of the XX and Heisenberg
critical spin chains the central charge is $c=\bar c=1$ and the model falls
into the free boson universality class. The critical Ising model
corresponds to a free fermion, thus $c=\bar c=\frac{1}{2}$. The
central charge of the theory is seen to play the role of a measure of
entanglement.  The vacuum of a theory of free bosons is more entangled
than the one corresponding to a theory of free fermions.  Scaling
of entanglement is just another manifestation of the ubiquitous
organizing principle orchestrated by conformal symmetry. The amount of
surprise (or entropy) obtained when in a given theory a new degree of freedom (a new spin in the block) is added must follow scaling as dictated by the representation of
conformal symmetry corresponding to that theory. Because the entropy
of the reduced density matrix of the ground state is not attached to
any particular operator, it is natural that the central charge is the
parameter in control of this measure of entanglement.

Due to the relation between entanglement and the central charge
a number of further connections appear. The central charge
quantifies quantum correlations as well as {\it e.g.} the trace anomaly.
A plausible interpretation can now be given in terms of entanglement. The more quantum correlated the vacuum is, the stronger the breaking
of conformal symmetry appears when a curved background is present.
More relevant is how the theorem on irreversibility
of renormalization group flows  translates to entanglement 
in a way that will be discussed shortly.


\subsection{Entanglement in higher dimensional systems}

Entanglement in spin chains obeys logarithmic scaling.  This
result was seen to emerge from conformal symmetry.  It is then
possible to apply similar arguments in higher dimensions and complete
them with standard arguments based on the Schmidt decomposition.

Let us consider a $d+1$ field dimensional theory at its critical point.  Its
ground state is a pure state. Following Srednicki~\cite{Sr93}, one
could now consider the reduced system on {\it e.g.}  a hypersphere
$S^{d-1}$ of radius $R$. The division into the interior $A$ and the
exterior $B$ of this imaginary hypersphere can be used to write the
ground state using the Schmidt decomposition in terms of pure states
$\vert \xi_i\ra, \vert\phi_i\ra$ associated to each region \be \ket{0} =\sum_i \sqrt{p_i} \vert \xi_i\ra_A \vert \phi_i\ra_B\ , \ee
where $p_i$ are positive numbers and the sum ranges up to the minimum
of the dimensions of the Hilbert spaces for A and B.  The standard
argument follows that both reduced density matrices share the same
eigenvalues $p_i$ and, thus, the same entropy. Yet, both systems only
share the hypersurface separating them, so that the leading term of
the entropy as any infrared cutoff is sent to infinity and the
ultraviolet cutoff $x_{uv}$ is sent to zero must scale as its ``area''
\be S_R =c_1~ \left( \frac{L}{x_{uv}} \right)^{d-1} \label{eq:d1} \ee where $c_1$ is a
known coefficient related to anomalies that we shall discuss
later. This leading scaling law for $x_{uv}\to 0$ also holds for massive
particles as checked by explicit calculations in free massive theories
\cite{kabat}.  The ``area'' law associated with the hypersurface is
understood as an effect coming from the loss of coherence between the
points at each side of the boundary separating the interior and
exterior parts of the universe.  It is then natural to expect that
microscopic condensed matter systems will follow the same law.

At variance with spin chains, the ultraviolet cutoff $x_{uv}$ gets now
mixed with the global coefficient $c_1$ and it is unclear how to
extract observable information from the latter. It has been shown that $c_1$
corresponds to the coefficient of the linear term in the curvature in
the effective action of a field theory in a non-trivial gravitational
background \cite{kabat}.  Then, $c_1$ equals 1/6 for scalar particles and
1/12 $2^{[d/2]}$ for Dirac fermions. More precisely, every fermionic
component contributes to $c_1$ as half a boson. In 1+1 dimensions, we
worked with spinless fermions, thus the relative factor of 2 between
1/6 for the critical Ising model and 1/3 for the XX and Heisenberg critical
chains. In dimension 3+1, the entropy of a system of a free Dirac
fermion will carry twice more entropy than a free boson since the
Dirac fermion is made of four components.

Entanglement is thus also connected through Eq. (\ref{eq:d1}) to the effective action of quantum field
theories on gravitational backgrounds.
 It is a remarkable fact that in $1+1$ dimensions the effective
action has a unique non-local form proportional
to the central charge $c$. When this non-local action
is expanded in powers of the curvature, all terms carry
the dependence in $c$. It follows that the trace anomaly, which
is a derivative of this effective action with respect to
the metric, is also proportional to the central charge. This is no
longer true in higher dimensions.  The effective action of a quantum
field theory defined on a gravitational background develops infinitely
many apparently unrelated structures. 
The entropy of entanglement seems to be related to
$c_1$ \cite{kabat}, the coefficient of the linear term in the curvature
$R$ (called $a_1$ in \cite{birrel}).  On the other hand, other
contributions to the trace anomaly giving rise to the Euler density
and to non-trivial two-point energy-momentum tensor correlators are
related to the structures that come quadratic with the curvature.
All the coefficients in the effective action seem to quantify
different aspects of entanglement.

A remarkable result concerning the saturation of entanglement in
non-critical systems can also be translated from quantum field theory
to spin chains. It has been proven \cite{kabat} that the leading order result when the ultraviolet cutoff is sent to infinity, Eq. (\ref{eq:d1}) is not
modified by adding a mass to the scalar field.  The leading behavior of
entanglement is not affected by going away from conformal symmetry
because the contribution to the entropy comes from the entanglement
between points near each side of the boundary (see also \cite{Ga01}). 
The scaling law in
terms of the hypersurface separating them is respected although a
finite correlation length is present.  Moreover, the subleading
corrections to the ``area'' law are known when $x_{uv}\ll \frac{1}{m}\ll
x_{ir}$, $x_{ir}$ being an infrared cutoff which defines a separation
of space in two regions in analogy to our previous $R$, 
\be
\label{mass1}
S_{m}-S_{m= 0}=
 \begin{cases} -\frac{x_{ir}m}{24 \pi}& d=2+1\cr
\frac{x_{ir}^2m^2}{96 \pi^2}\ln {m x_{uv}}& d=3+1\cr
\end{cases}
\ee 
where the subindex $m$ labels the massive theory.  The case
$d=1+1$ corresponds to the universality class we have been discussing
in systems of spin chains and yields the following result \cite{kabat}\footnote{We quote this equation from D. Kabat and M. J. Strassler works in spite of the fact that the logarithm is in the natural basis. Different definitions of entropy are given in different communities but the basis just defines the unit of measure.}
\be
\label{mass2}
S=-\frac{1}{6}\ln m x_{uv}\qquad x_{uv}\ll \frac{1}{m}\ll x_{ir} \ .\ee
This latter result is indeed observed in our computation (and will be presented more extensively in some other publication). For instance, the external magnetic field deformation of the Ising model $(\gamma=1, \lambda \neq 1)$, Fig.~(\ref{ising}), corresponds to an effective mass $m \sim |1-\lambda|$. 

The lack of saturation of entanglement in massive theories
have clear implications for the application of DMRG
techniques as well as other modifications of the Wilsonian block-spin
idea for condensed matter systems.
Entanglement always diverges (even off criticality)
in more than 1+1 dimensions and limits the
success of such an approach.  It seems conceptually preferable to
construct projections of the exact renormalization group based in
momentum space as the local potential approximation. This
method has proven quite powerful in scalar theories \cite{erg}
where it provides 
good approximations to critical exponents  in any number 
of dimensions and can even
detect triviality.

\subsection{Majorization and entanglement}
\label{subsec:majo_and_ent}

Let us return to quantum spin chains where 
entanglement pervades the system at conformal points. The absence of a
mass scale makes quantum correlations extend to long
distances. Consequently, the vacuum structure must describe this fact
and the entropy shows scaling as discussed above. It is arguable that
the entropy of the vacuum in the reduced system is just one out of
many possible measures of entanglement.  Then, other yet unexplored
measures of entanglement may bring further information about the
structure of the vacuum.

In order to investigate this issue, we can further exploit our
computations due to the fact that we have explicit results for the
density matrix of the ground state
in the Ising and XX models. The $2^L$ eigenvalues of $\rho_L$
correspond to the product of the $2$ eigenvalues $(1\pm \nu_m)/2$ of $L$ fermionic modes, Eq. (\ref{eq:rhoprod}),
 \be
\label{rhoeigen}
\lambda_{x_1x_2\cdots x_L} = \prod_{m=0}^{L-1}
\frac{1+(-1)^{x_m}\nu_m}{2},~~~x_m = 0,1~ \forall m.  
\ee 
It is
convenient to first visualize the typical shape of the eigenvalues
$(1 \pm \nu_m)/2$ of the fermionic modes and observe that most of the $\rho_L$ eigenvalues are almost zero or one and only a small
set will take intermediate values, bringing the main contribution to
the entropy. Each mode tends to remember its pure state origin.
\begin{center}
\begin{figure}[!ht]
\resizebox{!}{4.2cm}{\includegraphics{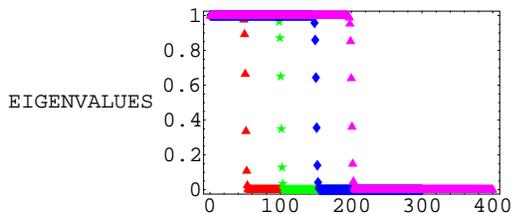}}
\caption{\label{maj1}Plot of the eigenvalues $\frac{1 \pm
\nu_m}{2}$ of the fermionic modes, Eq. (\ref{eq:rhoprod}), for the Ising model, Eq. (\ref{eq:Ising}), and $L=\{50$ (triangles), $100$
(stars), $150$ (diamonds), $200$ (triangles)$\}$. As $L$ increases new
modes are populated.}
\end{figure}
\end{center}
\begin{center}
\begin{figure}[!ht]
\resizebox{!}{4.2cm}{\includegraphics{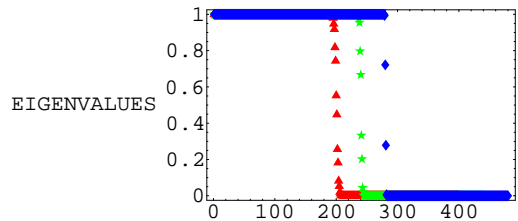}}
\caption{\label{maj2}Plot of the eigenvalues $\frac{1 \pm \nu_m}{2}$ of the fermionic modes, Eq. (\ref{eq:rhoprod}),
for the XX model as a function of the magnetic field $\lambda$, Eq. (\ref{eq:XX}), taking $L=200$ spins from
an infinite ring. Triangles display the eigenvalues for $\lambda=0.5$,
stars for $\lambda=0.999$ and diamonds the eigenvalues for
$\lambda=0.99999$. When $\lambda$ approaches 1, the ground state
becomes a product state and the modes take only the value one or
zero. The three plots should overlap each other, but we have artificially shifted each plot 40 points respect to the previous one.}
\end{figure}
\end{center}

The eigenvalues of $\rho_L$ form a
probability distribution. It is numerically easy verified that at
conformal points 
\be \rho_{L+2}\prec \rho_L\ ,\ee
 that is, the
probability distributions associated to the density matrices obey a
majorization relation. Due to the spin structure of the problem the
majorization takes place at two-spin steps. We have verified that
a similar result about majorization holds for the quantum Ising model at its critical
point.
\begin{center}
\begin{figure}[!ht]
\resizebox{!}{4.2cm}{\includegraphics{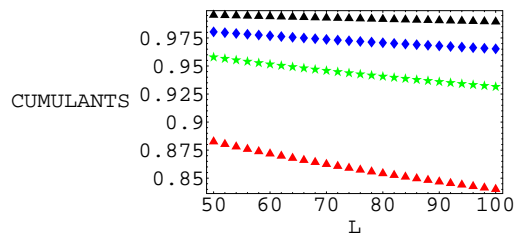}}
\caption{\label{maj3} Plot of the cumulants of the 8 (triangles),
16 (stars), 32 (diamonds) and 64 (triangles)
larger probabilities associated to the ground state $\rho_L$ as a
function of $L$ in steps of 2 in the XX model without magnetic field.}
\end{figure}
\end{center}

The increase of amount of surprise quantified by the entropy is rooted
in a deeper sense of ordering in the vacuum.  The ground state $\rho_L$ becomes
more and more disordered as dictated by a majorization arrow. Note
that this implies and increasing number of relations between the
eigenvalues of $\rho_L$ as $L$ goes to infinity. The vacuum
in a field theory is far more ordered than what the scaling of entropy 
hints at, in the sense that majorization provides a more strict definition of order than entropy.

In order to establish the role of  majorization and its relation
to conformal symmetry, further work is necessary. In particular,
marginal deformations need to be analyzed carefully. Later on we shall
argue that a different type of majorization seems to hold
along renormalization group flows. In that case, $L$ is kept fixed
but the parameters of the hamiltonian change as dictated
by renormalization group transformations. 

\subsection{Irreversibility of entanglement and the c-theorem}

Given the relation between entanglement and conformal symmetry, some
powerful quantum field theory results can be borrowed and translated
to quantum information parlance. It is of particular relevance the so
called Zamolodchikov's c-theorem \cite{zam86} that establishes that
the central charge of a unitary 1+1 dimensional theory always
decreases along renormalization group trajectories.  The existence of
a c-theorem in more than 1+1 dimensions has been a subject of a lot of
effort \cite{Ca88,cfl91,clv92,osbo,anse} and a proof has been proposed in
\cite{fl98}.  Zamolodchikov's theorem in 1+1 dimensions
can be proven using the
spectral densities following Ref. \cite{cfl91}. Consider the spectral
representation of the correlator for two energy-momentum tensors
(the energy-momentum operator 
 is defined for any quantum field theory
and corresponds to a descendant of the identity)
\be
\begin{split} 
&\langle 0\vert T_{\alpha\beta}(x) T_{\mu\nu}(0)
\vert 0 \ra =\\ &= \frac{\pi}{3}
\int d\lambda \, c(\lambda,\mu) \left( p_{\alpha} p_{\beta} - p^2
g_{\alpha\beta} \right) \left(p_{\mu} p_{\nu} - p^2 g_{\mu\nu} \right)
G(x,\lambda)
\end{split}
\ee 
where $\lambda$ is the spectral parameter (with dimensions of
mass), $c(\lambda,\mu)$ is the spectral function, which depends on
$\lambda$ and on the subtraction point $\mu$, and $G(x,\lambda)$ is
the free scalar propagator of a particle with mass $\lambda$. At a
fixed point of the renormalization group flow, the spectral function reduces to a delta distribution
$\left. c(\lambda,\mu)\right|_{cft}= c\ \delta(\lambda)$, where the
coefficient $c$ is a constant, reflecting the fact that all physical
intermediate states are massless.  The UV fixed point can be analyzed
taking $x\to 0$ and it follows that 
\be c_{UV}=\int d\lambda \,
c(\lambda,\mu) .  
\ee 
where $c_{UV}$ is the central charge of the
ultraviolet theory.  On the other hand, in the IR limit only massless
modes survive, so that the spectral function can in general be written
as 
\be c(\lambda,\mu)= c_{IR}\ \delta(\lambda) + c_{\rm
smooth}(\lambda,\mu)\ , \ee 
where the contribution of all massive
modes is contained in $c_{\rm smooth}(\lambda,\mu)$.  It thus
follows that 
\be c_{UV}=c_{IR}+\int d\lambda \, c_{\rm
smooth}(\lambda,\mu) , \ee 
where the second term on the r.h.s. is
necessarily $\mu$--independent.  Finally, unitarity guarantees that
$c_{\rm smooth}$ is positive, so \be c_{UV}\ge c_{IR}.  \ee This result
can be understood as a net decrease of degrees of freedom as weighted
by the central charge along renormalization group trajectories. Note
also these ideas match bosonization: in 1+1 dimensions two Majorana
fermions can be made into a boson. All proposed generalizations of the
central charge in more than 1+1 dimensions are based on the trace
anomaly and share the property that fermions weight more than
bosons. Exact bosonization is then not possible.

The c-theorem in 1+1 dimensions immediately implies that entanglement decreases along
renormalization group flows. Critical regions in any theory are given by the fixed points of the renormalization group transformations and, in 1+1 dimensions, they are characterized by the central charge of the model. As we have seen, the entanglement between a block of spins and the rest of the chain in a critical region is given by the central charge times the logarithm of the length of the block; and from the c-theorem, the renormalization flow of any unitary theory goes from an ultraviolet fixed point to an infrared fixed point, i.e., from a higher to a smaller central charge value. Then, the entropy of the reduced density matrix decreases along renormalization group
trajectories, 
\be S^{UV}_L > S^{IR}_L .  \ee
The decoupling of the massive modes in the spectrum at long distances
irreversibly reduces the amount of non-local quantum correlations in
the system. This result is only valid for unitary theories and is not
obvious because a renormalization group transformation in quantum
field theory is made of two steps: integration of high-energy modes
followed by rescaling. While the first step seems to imply
irreversibility, the second makes it obscure. Furthermore, the proper
construction of this result implies a careful treatment of
renormalization.  The role of unitarity is dominant since
counterexamples can be build where renormalization group flows form
limit cycles for non-unitary theories.  Entanglement loss and
unitarity are, thus, related.

There is an explicit result we have already presented that can be used
to illustrate the loss of entanglement along renormalization group
trajectories. Consider the non-critical massive boson. The departure
from the massless case corresponds to adding a relevant operator. The
flow will make the mass grow. The renormalization group trajectory is
driven by the increase of the mass. It must follow from the
(1+1)-dimensional c-theorem that for increasing mass, the entropy of the
system decreases.  This is indeed the case as verified in
Eq. (\ref{mass2}) 
\be 
 S_{L,m_1}-S_{L,m_2} =-\frac{1}{ 6} \ln \frac{m_1}{
m_2} < 0 \qquad m_1> m_2.  
\ee 
A perturbative c-theorem is also known
to hold in any dimension and, again, it can be checked on
Eq. (\ref{mass1}) 
\be
  S_{L,m}-S_{L,m=0}<0  \ .
\ee 
Furthermore the
decrease of the entropy is monotonous in $m$.

It is natural to try to go one step further and try to relate
irreversibility of the renormalization group to the majorization
properties that may structure the entanglement in the vacuum. Our
preliminary numerical results seem to indicate that 
\be
 \rho_L^{m=0}\prec \rho_L^{m} 
\ee 
in spin chains, in all the
analytical results shown in Eqs.(\ref{mass1}) and (\ref{mass2}) and in the
computation produced by Srednicki \cite{Sr93}. Although all the above
cases correspond to renormalization group flows driven by 
a simple massive deformation,
one then may speculate
that 
\be 
 \rho_L^{UV}\prec \rho_L^{IR} 
\ee
 and that majorization is
indeed underlying the irreversibility of renormalization group flows.
This fact remains a conjecture, nevertheless there are some numerical evidences that will be shown in a later publication.

\section*{Acknowledgments}

We thank A. Kitaev for very valuable input and R. Emparan for fruitful discussions. This work was supported by Spanish grants GC2001SGR-00065
and MCYT FPA2001-3598, by the National Science Foundation of USA under
grant EIA--0086038, and by the European Union under grant ISF1999-11053.


\appendix

\section{Spectrum of $H_{XY}$}
\label{app:app1}

The material that is shown in this appendix can be found in any standard text book that studies spin systems (see for instance \cite{sach99,hen01,chak01}). Nevertheless, we present it to make the article self-contained.
In this appendix, we determine the spectrum of the XY model with transverse magnetic field and open boundary conditions,
\bea
H_{XY} &=& -\frac{1}{2} \!\sum_{l=-\frac{N-1}{2}}^{\frac{N-1}{2}} 
\left( \frac{1\!+\!\gamma}{2} \sigma_l^x \sigma_{l+1}^x +
 \frac{1\!-\!\gamma}{2} \sigma_l^y \sigma_{l+1}^y \right) \nonumber\\
 &-& \frac{1}{2} \sum_{l=-\frac{N-1}{2}}^{\frac{N-1}{2}} \lambda\sigma_l^z,
\label{eq:XY3}
\eea
in the limiting case of an infinite chain, $N\rightarrow \infty$.
The spectrum of this model was originally computed by Katsura \cite{Kat}, generalizing the results of Lieb, Schultz and Mattis \cite{Ann} for the XY model without magnetic field.

\subsubsection{Jordan-Wigner Transformation}

The initial spin operators satisfy anticommutation rules at any given
site but follow commutation rules at separate sites. The non-local
Jordan-Wigner transformation maps these operators into fully
anticommuting spinless fermions defined by 
\be
\begin{split} 
&\Fa_l = \left( \prod_{m<l}\sigma_m^z \right) \frac{\sigma_l^x -
 i\sigma_l^y}{2} \\ 
&\{\Fa_l^{\dagger},\Fa_m\}=\delta_{lm} , ~~~\{\Fa_l,\Fa_m\} =0.
\end{split}
\ee
In terms of operators $\Fa$ the above Hamiltonian becomes
\be
\begin{split}
H_{XY}=&\frac{1}{2} \sum_{l=-\frac{N-1}{2}}^{\frac{N-1}{2}}
 [ \left(\Fa_{l+1}^\dagger \Fa_{l} + \Fa_l^\dagger \Fa_{l+1}\right)\\ 
& + \gamma \left(\Fa_l^\dagger \Fa_{l+1}^\dagger + \Fa_{l+1} \Fa_l
 \right)]
 - \lambda\!\!\! \sum_{l=-\frac{N-1}{2}}^{\frac{N-1}{2}}
 \!\!\Fa_l^{\dagger} \Fa_l .
\end{split}
\ee


\subsubsection{Fourier Transformation}

We can now exploit the (quasi) translational symmetry of the system by
introducing Fourier transformed fermionic operators 
\be
\Fd_k =\frac{1}{\sqrt{N}} \sum_{l=-\frac{N-1}{2}}^{\frac{N-1}{2}} 
\Fa_l e^{-i\frac{2\pi}{N}kl},
\ee 
$-(N-1)/2 \leq k \leq (N-1)/2$. Due to the fact that this transformation
 is unitary, the anticommutation relations remain valid 
\be 
\{\Fd^{\dagger}_k,\Fd_p\}=\delta_{kp}
\hspace{5ex} \forall \{k,p\}.  
\ee 
The Hamiltonian now takes an almost diagonal form, 
\be
\begin{split}
H=&\sum_{k=-(N-1)/2}^{(N-1)/2} (-\lambda+\cos \frac{2\pi k}{N}) \, 
\Fd^{\dagger}_k \Fd_k + \\
&\frac{i \gamma}{2} \sum_{k=-(N-1)/2}^{(N-1)/2} \sin\frac{2\pi k}{N} \, ( \Fd_k
\Fd_{-k} + \Fd^{\dagger}_k \Fd^{\dagger}_{-k}).
\end{split}
\ee
where an extra term, suppressed by $\frac{1}{N}$, should be present. 
 We shall though ignore it here since our results concern the limit
 $N\to \infty$.

\subsubsection{Bogoliubov Transformation}

A final unitary transformation is needed to cast the Hamiltonian into
a manifestly free particle theory. This so-called Bogoliubov
transformation can be expressed as 
\be
\begin{split}
\Fb^{\dagger}_k =u_k \, \Fd^{\dagger}_k + i v_k \, 
\Fd_{-k} \\ \Fb_k = u_k \, \Fd_k - i
v_k \, \Fd^{\dagger}_{-k},
\end{split}
\ee 
where $u_k = \cos{\theta_k/2},~ v_k = \sin{\theta_k/2}$ for
\be 
\cos{\theta_k} = \frac{-\lambda+\cos{\frac{2\pi}{N}k}}
{\sqrt{(\lambda-\cos{\frac{2\pi k}{N}})^2+\gamma^2 \sin^2{\frac{2\pi k}{N}}}}.
\ee
Again, due to unitarity of the Bogoliubov transformation the operators
$\{\Fb_k\}$ follow the usual anticommutation relation 
\be
\{\Fb^{\dagger}_k,\Fb_p\}=\delta_{kp} \hspace{5ex} \forall \{k,p\}.  
\ee
Finally, the Hamiltonian takes a diagonal form 
\be
H=\sum_{k=-(N-1)/2}^{(N-1)/2} \tilde{\Lambda}_k
 \, \Fb^{\dagger}_k \Fb_k,
\ee
where
\be
\tilde{\Lambda}_k \equiv \sqrt{\left(\lambda-\cos{\frac{2\pi
k}{N}}\right)^2+\gamma^2 \sin^2{\frac{2\pi k}{N}}}.
\ee
The thermodynamical limit is obtained by defining $ \phi =2\pi k/N$
and taking the $N\to \infty$ limit \be H = \int^{\pi}_{-\pi}\frac{{\rm
d}\phi}{2\pi}\ \Lambda_\phi \Fb^{\dagger}_\phi \Fb_\phi, \ee with \be
\label{energy}
\Lambda^2_\phi = (\lambda-\cos{\phi})^2 + \gamma^2 \sin^2{\phi}.  \ee


\section{Correlation matrix for the XY model}
\label{app:app2}

In this section we show how to obtain the correlation 
matrix $\ev{\Ma_m\Ma_n}$ of the ground state of the Hamiltonian 
\bea
H_{XY} &=& \frac{i}{2} \sum_{l=-\frac{N-1}{2}}^{\frac{N-1}{2}} 
\left(~\frac{1+\gamma}{2}~\Ma_{2l}\Ma_{2l+1}
-  \frac{1-\gamma}{2}~\Ma_{2l-1}\Ma_{2l+2}~ \right) \nonumber \\
&+& \frac{i}{2} \sum_{l=-\frac{N-1}{2}}^{\frac{N-1}{2}} \lambda \Ma_{2l-1}\Ma_{2l}~,
\label{eq:amajXY}
\eea
where the Majorana operators $\Ma$ fulfill
\be
\Ma_m^\dagger = \Ma_m,~~~~~~\{a_m,a_n\}=2\delta_{mn},
\ee
$-N \leq m,n \leq N-1$. 

These correlators were originally computed by Lieb, Schultz and Mattis \cite{Ann} for the XY model without magnetic field and by Barouch and McCoy \cite{Bar} for the XY model with magnetic field.

We start by diagonalizing the above Hamiltonian. This 
can be achieved by means of two canonical transformations. 
Let us define $2N$ auxiliary Majorana operators $\Md$ and $\Me$,
\bea
\left[ \begin{array}{cc}
\Md_{2k-1} \\
\Md_{2k}
\end{array} \right]
&=& \sqrt{\frac{2}{N}} \sum_{l=-\frac{N-1}{2}}^{\frac{N-1}{2}}
 \cos \frac{2\pi kl}{N} ~~
\left[ \begin{array}{cc}
\Ma_{2l-1} \\
\Ma_{2l}
\end{array} \right],\\
\left[ \begin{array}{cc}
\Me_{2k-1} \\
\Me_{2k}
\end{array} \right]
&=& \sqrt{\frac{2}{N}} \sum_{l=-\frac{N-1}{2}}^{\frac{N-1}{2}}
 \sin \frac{2\pi kl}{N} ~~
\left[ \begin{array}{cc}
\Ma_{2l-1} \\
\Ma_{2l}
\end{array} \right],
\eea
$0\leq k \le N/2$, that take the Hamiltonian into sum of Hamiltonians $H_k$,
\be
H_{XY} = \sum_{k=0}^{N/2} H_k,
\ee
where
\be
H_k = \frac{i\tilde{\Lambda}_{k}}{4}  
\left[\begin{array}{c} \Md_{2k-1} \\ \Me_{2k-1} \\ \Md_{2k} 
\\ \Me_{2k} \end{array}\right]^T
\left[\begin{array}{cccc}
0 & 0 & c_k & -s_k \\
0 & 0 & s_k & c_k \\
-c_k & -s_k & 0 & 0 \\
s_k & -c_k & 0 & 0 
\end{array}\right]
\left[\begin{array}{c} \Md_{2k-1} \\ \Me_{2k-1} \\ \Md_{2k} 
\\ \Me_{2k} \end{array}\right],
\ee
$c_k \equiv \cos \theta_k$, $s_k \equiv \sin \theta_k$. 
The diagonalization is completed by a second transformation
 that acts independently for each value of $k$, according to
\be
\left[\begin{array}{c} \Mb_{-2k-1} \\ \Mb_{-2k} \\ \Mb_{2k-1} 
\\ \Mb_{2k} \end{array}\right] = \frac{1}{\sqrt{2}}
\left[\begin{array}{cccc}
u_k & v_k & u_k & -v_k \\
u_k & v_k & -u_k & v_k \\
v_k & -u_k & v_k & u_k \\
-v_k & u_k & v_k & u_k 
\end{array}\right]
\left[\begin{array}{c} \Md_{2k-1} \\ \Me_{2k-1} \\ \Md_{2k}
 \\ \Me_{2k} \end{array}\right].
\ee
In terms of the $2N$ Majorana operators $\Mb_p$, $-N 
\leq p \leq N-1$, the Hamiltonian is finally written as
\be
H_{XY} = \frac{i}{4}\sum_{k=-\frac{N-1}{2}}^{\frac{N-1}{2}}
 \tilde{\Lambda}_k (\Mb_{2k-1}\Mb_{2k}-\Mb_{2k}\Mb_{2k-1}),
\ee
as we wanted to show. 

The above two orthogonal transformations define $W\in SO(2N)$, where
\be
\Mb_p = \sum_{m=-N}^{N-1} W_{pm} \Ma_m, ~~~-N+1 \leq p \leq N.
\ee
Then, coefficients $g_l$ in Eq. (\ref{eq:GammaA}) are obtained through
\be
\Gamma^A = W^{T}\Gamma^D W,
\ee
where
\be
\Gamma^B = \bigoplus_{k=-\frac{N-1}{2}}^{\frac{N-1}{2}} \left[
\begin{array}{rc}
0 & 1  \\
- 1 &0
\end{array}
\right].
\ee


\section{Analytical evaluation of the correlation matrix for the XY model}
\label{app:app3}

In this appendix we present an analytical expression 
for $g_l$, Eq. (\ref{eq:g}), for five particular cases of the XY chain.

\subsubsection{Ferromagnetic limit}
The ferromagnetic limit corresponds to $\lambda \to \infty$. In this
case, it is easy to see that 
\be g_l =
\delta_{l0} \hspace{5ex} \forall \, \gamma. 
\ee

\subsubsection{Ising model}

For an Ising model, $\gamma=1$,  with magnetic field $\lambda$ we have
\be 
g_l = \frac{1}{2\pi}\int^{\pi}_{\pi} d\phi
e^{-il\phi} \frac{e^{-i\phi}-\lambda}{\sqrt{1-\lambda
e^{i\phi}}\sqrt{1-\lambda e^{-i\phi}} }.  
\ee 
For values of $\lambda\in[-1,+1]$, 
\be 
(1-\lambda
e^{i\phi})^{-\1/2}=\sum^{\infty}_{m=0}
\frac{(2m-1)!!}{(2m)!!}\lambda^m e^{i\phi m}, 
\ee 
where 
\be 
(2m)!!=
2^m m! ~~~~~~~~ (2m-1)!!= \frac{(2m)!}{2^m m!}, 
\ee 
obtaining the equation 
\be 
\begin{split}
&g_l=\\
&\begin{cases}
\sum^{\infty}_{m=0} \left( \frac{(2(m+l+1))!}{(2^{m+l+1}(m+l+1)!)^2} -
\frac{(2(m+l))!}{(2^{m+l}(m+l)!)^2}
\right)\\ \frac{(2m-1)!!}{(2m)!!}\lambda^{2m+l+1}~~~~~ l \ge 0,\\ \left(
\sum^{\infty}_{m=-l-1} \frac{(2(m+l+1))!}{(2^{m+l+1}(m+l+1)!)^2}-
\sum^{\infty}_{m=-l} \frac{(2(m+l))!}{(2^{m+l}(m+l)!)^2}
\right)\\ \frac{(2m-1)!!}{(2m)!!}\lambda^{2m+l+1}~~~~~ l \le 0.\\
\end{cases}
\end{split}
\ee

For the limits $\lambda=0$ or $\lambda=1$ these equations reduce to

\paragraph{Ising model without magnetic field.-}
The Ising model is recovered for $\gamma = 1$. When no magnetic field
is present, $\lambda = 0$ and $g_l$
reduces to \be g_l = \delta_{-l1}. \ee

\paragraph{Ising model in a critical magnetic field.-}
For the Ising model, $\gamma = 1$, and when the magnetic field is in
the critical point $\lambda = 1$, the initial expression for $g_l$
transforms to, 
\be
\begin{split}
g_l&=\int^{\pi}_0 \frac{d\phi}{\pi} \frac{(-1+\cos{\phi})\cos{\phi l}-
\sin{\phi}\sin{\phi l} }{ \sqrt{(1-\cos{\phi})^2+ \sin^2{\phi}} }\\ &=
\int^{\pi}_0 - \frac{d\phi}{\pi} \sin{\left(l+\1/2 \right)\phi} =
\frac{-1}{\pi \left(l+\1/2 \right)}.
\end{split}
\ee

\subsubsection{XX model with magnetic field}
The XX model corresponds to $\gamma = 0$ with a magnetic field in the
range given by $ \lambda \in [-1,1]$.  Making a small transformation
in the general expression, we get 
\be
\begin{split}
g_l &= \int^{\pi}_0 \frac{d\phi}{\pi}
\frac{-\lambda+\cos{\phi}}{|\lambda-\cos{\phi}|} \cos{l\phi}\\ &=
\frac{1}{\pi} \left(\int^{\phi_c}_0 \cos{l\phi} \, d\phi -
\int^{\pi}_{\phi_c} \cos{l\phi} \, d\phi \right) \\ 
&=\begin{cases} 
\frac{2}{l\pi}
\sin{l\phi_c} & l \neq 0\\
\frac{2\phi_c}{\pi}-1 & l = 0,
\end{cases}
\end{split}
\ee 
where $\phi_c = \arccos{(\lambda)}$.
In the particular subcase of the XX model without magnetic field,
$\lambda=0$, we obtain \be g_l =\frac{2}{l\pi} \sin{\frac{l\pi}{2}},
\ee which is equivalent to \be
\begin{split}
g_l & = 0, \hspace{5ex} l \in even \\ g_l & = \frac{2}{l\pi}
(-1)^{(l-1)/2}, \hspace{5ex} l \in odd.
\end{split}
\ee

\subsubsection{The XY model with critical magnetic field}
For any anisotropy $|\gamma| \le 1$ and critical magnetic field,
$\lambda =1$, the general expression is recast into,

\be
\begin{split}
g_l =& \int^{\pi}_0 \frac{d\phi}{\pi}
\frac{(\cos{\phi}-1)\cos{l\phi}-\gamma\sin{\phi}
\sin{l\phi}}{\sqrt{(1-\cos{\phi})^2+\gamma^2\sin^2{\phi}}}\\
=&-\frac{\gamma+1}{2\pi} \int^{\pi}_0 d\phi
\frac{\sin{(l+\1/2)\phi}}{\sqrt{\sin^2{\phi/2}+\gamma^2 \cos^2{\phi/2}}}\\
&-\frac{\gamma-1}{2\pi} \int^{\pi}_0 d\phi
\frac{\sin{(l-\1/2)\phi}}{\sqrt{\sin^2{\phi/2}+\gamma^2 \cos^2{\phi/2}}}
\end{split}
\ee

\subsubsection{The XY model without magnetic field}
Finally, following reference~\cite{Ann} the limit of the XY model
without magnetic field corresponds to $g_l$ expressed as follows: \be
\begin{split}
g_l = &- \left(\frac{1+\gamma}{2} L_{l+1} + \frac{1-\gamma}{2} L_{l-1}
\right)
\hspace{5ex} l \in odd \\ g_l = & 0 \hspace{5ex} l \in even,
\end{split}
\ee where, \be L_l = \frac{2}{\pi} \int^{\pi/2}_0 d\phi
\frac{\cos{\phi l}}{\sqrt{\cos^2{\phi}+\gamma^2 \sin^2{\phi}}}.  \ee A
series expansion of this integral is given by: \be
\begin{split}
&L_l(\gamma) = (-1)^{l/2} \frac{2}{1+\gamma} \\ 
&\left( h_0 h_{l/2} -
\frac{\ln{1-\alpha^2}}{\pi} -\sum_{r=1}^{\infty} \alpha^{2r} \left(
\frac{1}{r\pi}-h_r h_{r+(l/2)} \right) \right)
\end{split}
\ee 
where, 
\be h_l = 2^{2l} \begin{pmatrix} 2l\\l\end{pmatrix}
\hspace{5ex} \alpha= \frac{1-\gamma}{1+\gamma}.  
\ee
\vfil

\eject


\begin{thebibliography}{99}

\bibitem{BCS57} J. Bardeen, L. N. Cooper and J. R. Schrieffer,
Phys. Rev. {\bf 108}, 1175 (1957).

\bibitem{La83} R. B. Laughlin, Phys. Rev. Lett. {\bf 50}, 1395 (1983).

\bibitem{sach99} S. Sachdev, {\em Quantum Phase Transitions}, Cambridge
Univ. Press (1999).

\bibitem{BeDi00} C. H. Bennett and D. P. DiVincenzo, Nature {\bf 404},
247 (2000).

\bibitem{book} M. A. Nielsen and I. L. Chuang, {\em Quantum computation
and quantum information}, Cambridge Univ. Press. (2000).

\bibitem{FortPhys} Special issue on experimental proposals for quantum
  computation, Fortschr. Phys. {\bf 48}, No. 9--11 (2000).

\bibitem{pres01} J. Preskill, J. Mod. Opt. {\bf 47}, 127
(2000),{\em quant-ph/9904022}.

\bibitem{gvprep} G. Vidal, {\em in preparation}.

\bibitem{vid02} G. Vidal, J. I. Latorre, E. Rico and A. Kitaev,
to appear in Phys. Rev. Lett., \emph{quant-ph/0211074}.


\bibitem{NiTe} M. A. Nielsen, PhD Thesis, University of New Mexico, New Mexico, USA (1998), {\em quant-ph/0011036}.

\bibitem{W00} W. K. Wootters, {\em quant-ph/0001114}.

\bibitem{OW01} K. M. O'Connor and W. K. Wootters, Phys. Rev. A {\bf 63}, 052302 (2001), {\em quant-ph/0009041}.

\bibitem{ved01} M. C. Arnesen, S. Bose and V. Vedral, Phys. Rev. Lett. {\bf 87}, 017901 (2001), {\em quant-ph/0009060}.

\bibitem{ved02} D. Gunlycke, S. Bose, V.M. Kendon, V. Vedral,
Phys. Rev. A {\bf 64}, 042302 (2001), {\em quant-ph/0102137}.

\bibitem{WFS01} X. Wang, H. Fu and A. I. Solomon, J. Phys. A: Math. Gen. 34 (50), 11307-11320 (2001), {\em quant-ph/0105075}.

\bibitem{WaZa02} X. Wang and P. Zanardi, Phys. Lett. A 301 (1-2), 1 (2002), {\em quant-ph/0202108}.

\bibitem{Wa02} X. Wang, Phys. Rev. A 66, 034302 (2002), {\em quant-ph/0203141}.

\bibitem{rest} P. Zanardi and X. Wang, J. Phys. A {\bf 35}, 7947 (2002), {\em quant-ph/0201028}; Yu Shi, {\em quant-ph/0204058} and {\em cond-mat/0205272}; K. Audenaert, J. Eisert, M.B. Plenio and R.F. Werner, Phys. Rev. A 66, 042327 (2002), {\em quant-ph/0205025}; M. A. Mart\'\i n-Delgado, {\em quant-ph/0207026}; J. Schliemann, {\em quant-ph/0212114}.


\bibitem{Os02} A. Osterloh, L. Amico, G. Falci and R. Fazio, Nature
{\bf 416}, 608 (2002), \emph{quant-ph/0202029}.

\bibitem{OsNi02} T. J. Osborne and M. A. Nielsen, Phys. Rev. A {\bf
66}, 032110 (2002), \emph{quant-ph/0202162}.


\bibitem{Ann} E. Lieb, T. Schultz and D. Mattis, Annals of Phys. {\bf
16} 407 (1961).

\bibitem{Kat} S. Katsura, Phys. Rev. {\bf 127} 1508 (1962)

\bibitem{Bar} E. Barouch and B. McCoy, Phys. Rev. A {\bf 3} 786 (1971)

\bibitem{bet01} H. Bethe, Z. Phys. {\bf 71}, 205 (1931).


\bibitem{Sr93} M. Srednicki, Phys. Rev. Lett. {\bf 71}, 666 (1993),
\emph{hep-th/9303048}.

\bibitem{Ca94} C. G. Callan and F. Wilczek, Phys. Lett.B {\bf 333}
(1994) 55, \emph{hep-th/9401072}.

\bibitem{Fi94} T. M. Fiola, J. Preskill, A. Strominger and
S. P. Trivedi, Phys. Rev. D {\bf 50} (1994) 3987,
\emph{hep-th/9403137}.

\bibitem{HoLaWi} C. Holzhey, F. Larsen and F. Wilczek, Nucl.Phys.B
{\bf 424} (1994) 443, \emph{hep-th/9403108}.

\bibitem{white} S. R. White, Phys. Rev. B {\bf 48}, 10345 (1993).

\bibitem{RO99} S. Rommer and S. \"Ostlund, ``Density matrix
renormalization'', Dresden, 1998 (Springer, Berlin, 1999), pp 67-89.

\bibitem{zam86} A. B. Zamolodchikov, JETP Lett. {\bf 43} (1986) 730.


\bibitem{benn01} C. H. Bennett, H. J. Bernstein, S. Popescu and
B. Schumacher, Phys. Rev. A 53, 2046 (1996), \emph{quant-ph/9511030}.

\bibitem{qic} Quan. Inf. Comp. {\bf 1} (2001).

\bibitem{Nie} M. A. Nielsen, Phys. Rev. Lett. {\bf 83}, 436 (1999),
\emph{quant-ph/9811053}.

\bibitem{mono} G. Vidal, J. Mod. Opt. {\bf 47}, 355 (2000),
\emph{quant-ph/9807077}. G. Vidal, Phys. Rev. Lett. {\bf 83}, 1046 (1999).
 


\bibitem{Dur} W. D\" ur, G. Vidal, J. I. Cirac, Phys. Rev. A {\bf 62},
062314 (2000), \emph{quant-ph/0005115}.

 
\bibitem{Vid03} G. Vidal, {\em quant-ph/0301063}.

\bibitem{woo01} W. K.Wootters, Phys. Rev. Lett. {\bf 80}, 2245
(1998),\emph{quant-ph/9709029}.

\bibitem{Woo03} V. Coffman, J. Kundu, W. K. Wootters, Phys. Rev. A {\bf 61} 052306 (2000), {\em quant-ph/9907047}.


\bibitem{petz} M. Ohya and D. Petz, {\em Quantum entropy and its use} (Springer-Verlag Berlin, 1993).

\bibitem{lie01} E. H. Lieb and M. B. Ruskai, J. Math. Phys., {\bf 14}
1938 (1973)

\bibitem{ste01} P. Stelmachovic and V. Buzek, presented at a quantum
information conference in Gdansk (July 2001) and in San Feliu (March
2002).

\bibitem{Bh96} R. Bhatia, {\em Matrix Analisis}, Graduate Texts in
Mathematics vol. 169, Springer-Verlag (1996).

\bibitem{gui02} M. A. Nielsen and G. Vidal, Quant. Inf. and Comp. {\bf
1}, 76 (2001).

\bibitem{niel01} T. J. Osborne and M. A. Nielsen, Quant. Inf. Proc. {\bf
1}, 45 (2002), {\em quant-ph/0109024}.

\bibitem{Pes} I. Peschel, M. Kaulke and  O. Legeza, Ann. Physik (Leipzig) {\bf 8} (1999) 153, {\em cond-mat/9810174}; M. C. Chung and I. Peschel Phys. Rev. B {\bf 64} 064412 (2001), {\em cond-mat/0103301}; I. Peschel, J. Phys. A: Math. Gen. {\bf 36} L205 (2003), {\em cond-mat/0212631}.

\bibitem{hen01}P. Christe and M. Henkel, {\em Introduction to conformal
invariance and its applications to critical phenomena}, 
Ed.~Springer-Verlag m.16.

\bibitem{chak01} B. K. Chakrabarti, A. Dutta and P. Sen, {\em Quantum
Ising phases and Transitions in Transverse Ising Models},
Ed. Spinger m.41.

\bibitem{kit} A. Kitaev, {\em cond-mat/0010440}.

\bibitem{orb01} R. Orbach, Phys. Rev. {\bf 112}, 309 (1958).

\bibitem{yan01} C. N. Yang and C. P. Yang, Phys. Rev. {\bf 150}, 321
(1966).

\bibitem{bon01} J. C. Bonner and M. E. Fisher, Phys. Rev. {\bf 135},
A640 (1964).

\bibitem{aff01} I. Affleck., J. Phys. A {\bf 31}, 4573 (1998),
\emph{cond-mat/9802045}.

\bibitem{Be94} J. D. Bekenstein 
\emph{gr-qc/9409015}.

\bibitem{cft} Paul Ginsparg, {\em Applied conformal field theory}
\newblock Les Houches Summer School 1988, pp 1-168.

\bibitem{kabat} D. Kabat and M. J. Strassler,
Phys. Lett. B {\bf 329},46 (1994), \emph{hep-th/9401125}; D. Kabat, Nucl. Phys. B{\bf 453}, 281 (1995), \emph{hep-th/9503016}.

\bibitem{birrel} N. D. Birrel and P. C. W. Davies, {\em Quantum fields in curved space}, Cambridge University Press (1982).

\bibitem{Ga01} J. Gaite, Mod. Phys. Lett. {\bf A16}, 1109 (2001), \emph{cond-mat/0106049}.

\bibitem{erg} T. R. Morris, Phys. Lett. B {\bf 329}, 241 (1994), \emph{hep-th/9403340};
J. Berges, N. Tetradis and C. Wetterich, Phys. Rept. {\bf 363}, 223 (2002), hep-ph/0005122;  
R. D. Ball et al. Phys. Lett. B {\bf 347}, 80 (1995), \emph{hep-th/9411122}.

\bibitem{Ca88} J. L. Cardy, Phys. Lett. B {\bf 215}, 749 (1988).

\bibitem{cfl91} A. Cappelli, D. Friedan and J.I. Latorre,
Nucl. Phys. B {\bf 352} (1991) 616.

\bibitem{clv92} A. Cappelli, J. I. Latorre and X. Vilas\'\i s-Cardona,
 Nucl. Phys. B {\bf 376}, 510
(1992), \emph{hep-th/9109041}.

\bibitem{osbo} H. Osborn and G.M. Shore, Nucl. Phys. B {\bf 571}, 287
(2000), \emph{hep-th/9909043}.

\bibitem{anse} D. Anselmi, Nucl. Phys. B {\bf 567}, 331 (2000),
 \emph{hep-th/9905005}.

\bibitem{fl98} S. Forte and J. I. Latorre, Nucl. Phys. B {\bf 535}
(1998) 709, \emph{hep-th/9805015}.


\end{thebibliography}
\end{document}